\documentclass[twocolumn,twocolappendix]{aastex7}
\usepackage{hyperref}

\newcommand{\ngc}{NGC\,1068}
\newcommand{\nh}{{\rm N}_{\rm H}}
\newcommand{\cms}{{\rm cm}^{-2}}
\newcommand{\cmc}{{\rm cm}^{-3}}

\newcommand\cdof{{\rm C/{\rm dof}}}
\newcommand{\chandra}{\textit{Chandra X-ray Observatory}}
\newcommand{\Chandra}{\textit{Chandra}}
\newcommand{\kms}{{\rm km\,s}^{-1}}
\newcommand{\fe}{Fe\,{\sc xxii}}
\newcommand\kev{{\rm\thinspace keV}}
\newcommand\erg{{\rm\thinspace erg}}
\newcommand\ergps{\hbox{${\rm erg\,s}^{-1}\,$}}
\newcommand\ergpcms{\hbox{${\rm erg\,cm\,s}^{-1}\,$}}
\begin{document}
\newcommand\Msun{\hbox{$\rm\thinspace M_{\odot}$}}
\title{On the \fe\ Emission in the X-ray spectrum of  \ngc}
\author[0009-0007-5987-0405]{M. Z. Buhariwalla}
\affiliation{Department of Astronomy and Physics, Saint Mary's University, 923 Robie Street, Halifax, NS B3H 3C3, Canada}
\email[show]{Margaret.Buhariwalla@smu.ca}
\author[0000-0003-2869-7682]{J. M. Miller}
\affiliation{Department of Astronomy, University of Michigan, Ann Arbor, MI 48109, USA}
\email{jonmm@umich.edu}
\author[0009-0006-4968-7108]{L. C. Gallo}
\affiliation{Department of Astronomy and Physics, Saint Mary's University, 923 Robie Street, Halifax, NS B3H 3C3, Canada}
\email{Luigi.Gallo@smu.ca}
\author[0000-0001-7557-9713]{J. Mao}
\affiliation{Department of Astronomy, Tsinghua University,  Beijing, 100084 China}
\email{jmao@tsinghua.edu.cn}
\author[0000-0002-7868-1622]{J. Raymond}
\affiliation{Harvard-Smithsonian Center for Astrophysics,  Cambridge, MA, USA}
\email{jraymond@cfa.harvard.edu}
\author[0000-0002-5779-6906]{T. Kallman}
\affiliation{NASA Goddard Space Flight Center, Greenbelt, MD 20771, USA}
\email{timothy.r.kallman@nasa.gov}

\begin{abstract}
The \fe\ doublet has been previously used to determine the density of collisionally ionized emission from magnetic cataclysmic variable stars. We test how this diagnostic doublet behaves for a photoionized plasma with an active galactic nucleus (AGN) spectral energy distribution (SED). 
We use the photoionized plasma code {\sc pion} and $\sim440$\,ks of archival Chandra HETG for the well-known Seyfert 2 galaxy \ngc\ to test the behaviour of the \fe\ doublet in the context of an AGN. This marks the first time these data have been examined with {\sc pion}. 
We find that in a photoionized plasma, the \fe\ doublet is dependent on the density, ionization state, and SED used. Thus, this density diagnostic remains model-dependent. In the context of \ngc\ the doublet predicts an emission region $\sim100\,r_g$ from the central black hole. This would require a direct line of sight to the central engine, which is at odds with the Seyfert 2 nature of this source. In practice, these results highlight the complexities and challenges of applying photoionized models. With these data, we cannot exclude the possibility of a direct line of sight to the central engine of \ngc, but we cannot confirm it. Future observations with instruments such as \textit{Athena} are needed to explore the \fe\ doublet further.  
\end{abstract}

\section{Introduction} 
\label{sec:intro}
The effects of ionized plasma are commonly seen in the soft X-ray emission of Active Galactic Nuclei (AGN). These effects appear either as photoionized emission lines \citep{Kinkhabwala+2002, Kallman+2014, Grafton+2021} or absorption from ionized plasmas called warm absorbers \citep[WA,][]{Blustin+2005, Tombesi+2013, Gallo+2023}. More than half of all unobscured ($\nh< 2\times 10^{22}\,\cms$) Seyfert galaxies are expected to host warm absorbers \citep{Laha+2014}. 
The properties of these warm absorbers and emitters can be inferred using line diagnostics ratios. In emission, the most commonly used are the G and R ratios of He-like ions \citep[e.g.][]{Porquet+2000, Kinkhabwala+2002}, while a variety of diagnostic ratios have been explored in absorption \citep[see][]{Mao+2017}.

There is the potential for a new diagnostic ratio utilizing doublet emission from \fe\ to probe the density of the emitting plasma. These lines originate from highly ionized B-like Fe, when an electron transitions from $2s^2 3d^1 \,{ }^2\mathrm{D}_{3/2} \rightarrow  2s^2 2p^1 \, { }^2\mathrm{P}_{1/2}$ producing an emission line at $11.77\,$\AA, a second emission line from \fe\ found at $11.92\,$\AA\ ($2s^2 3d^1 \,{ }^2\mathrm{D}_{5/2}\rightarrow  2s^2 2p^1 \, { }^2\mathrm{P}_{3/2}$)  accompanies the emission. Together, the ratio of these lines forms a density diagnostic sensitive between $n\sim10^{12}-10^{16}\,\cmc$ \citep{Mauche+2003, Mauche+2004, Liang+2008}. 

 The original \fe\ diagnostic curves were used to explore the density of magnetic cataclysmic variables, where the emission region is assumed to be either a completely collisionally ionized plasma or a combination of collisionally and photoionized plasmas \citep[see][]{Mauche+2003, Mauche+2004, Chen+2004, Liang+2008}. Photoionized absorption from \fe\ has been studied for AGN and X-ray binaries \citep[see][]{ King+2012, Miller+2014, Mao+2017, Ogorzalek+2022,Tomaru+2023}. Thus far, this diagnostic has not been applied to a photoionized emission spectrum of an AGN.  
 
\ngc\ is a Seyfert 2 galaxy located at a redshift $z=0.00379$ \citep{Huchra+1999} and observed through a Galactic column density of $\nh=3.22\times10^{20}\, \cms$ \citep{Willingale+2013}. This galaxy has been extensively observed using the High Energy Transmission Grating \citep[HETG,][]{Canizares+2000} aboard \chandra\  \citep[][]{Weisskopf+2000}. \cite{ Ogle+2003} reported the first observation from 2000,  where strong evidence of photoionized emission in the form of He-like triplets and radiative recombination continuum (RRC) was identified \citep[e.g.][]{Porquet+2000, Kinkhabwala+2002}. 
Evidence of outflowing winds was reported based on this first observation. Subsequent observations of \ngc\ were obtained and modelled with six outflowing winds \citep{Kallman+2014}. The ionization of these winds was fixed to be log\,$(\xi/\ergpcms)=1,\, 1.8,\, 2.6$ where $\xi=L/r^2n$ ($L$ is the ionizing luminosity between $1-1000$\,Ry, $r$ is the distance between the ionizing source and the emission region, and $n$ is the density of the emission region,  \citealt{Tarter+1969}). These winds outflow at $\sim-450\,\kms$, and they account for most of the emission lines observed in \ngc. The photoionized wind model found by \cite{Kallman+2014} does not account for the emission line features seen at $\sim 11.5-12\,$\AA. Thus, we conclude that not all emission components have been accounted for by these models. 
The  \fe\ line  at $11.77\,$\AA\ was identified in \ngc\ in the HETG \citep{Ogle+2003}, and RGS \citep{Grafton+2021} data. This makes \ngc\ a strong candidate to test the \fe\ diagnostic as a tool to find the density of emission regions in AGN. 

In this work, we explore the \fe\ density diagnostic and apply it to the HETG spectrum of \ngc. The Chapter is organized as follows: Section \ref{sec:data} outlines the data used in this work; Section \ref{sec:line} applies the literature line diagnostic; Section \ref{sec:spec} contains the spectral fitting results, and we explore the photoionized line diagnostic in Section \ref{sec:diagnostic}. We discuss the implications of our results in Section \ref{sec:discussion}, and summarize our conclusions in Section \ref{sec:conclusion}.

\section{Data} 
\label{sec:data}
\begin{table}
	\begin{tabular}{c c c c}
	\hline
(1)    & (2)            & (3)       \\
Obs ID &   Start date   &  Exposure \\
       & 	(yyyy-mm-dd) & [s]      \\     
	\hline
332    & 2000-12-04     & 45708   \\
9148   & 2008-12-05     & 80247   \\
9149   & 2008-11-19     & 89369   \\
9150   & 2008-11-27     & 41087   \\
10815  & 2008-11-20     & 19074   \\
10816  & 2008-11-18     & 16167   \\
10817  & 2008-11-22     & 33194   \\
10823  & 2008-11-25     & 34547   \\
10829  & 2008-11-30     & 39580   \\
10830  & 2008-12-03     & 43957   \\
\hline	
\end{tabular}
\caption
{Observational record of \ngc\ for \chandra\ HETG data. Column (1) The \chandra\ observational ID for each observation. Column (2) The start date of each observation, and Column (3) the exposure time of each observation.  }
\label{tab:obs}
\end{table}
We utilize $\sim443\,{\rm ks}$ of archival \Chandra\ data and Table \ref{tab:obs} lists the observations used. The HETG data were retrieved from {\sc TGCat}\footnote{\url{https://tgcat.mit.edu/}} \citep{Huenemoerder+2011}. The collection of observations has been archived with Chandra Data Collection (CDC) \href{https://doi.org/10.25574/cdc.466}{doi:10.25574/cdc.466}. The individual observations were summed using the {\sc ciao v.4.16.0} \citep{Fruscione+2006, Burke+2023} script {\sc combine\_grating\_spectra}. This script splits the pha2 files from each observation into their respective grating and order spectra. It then sums all the first-order (positive and negative) grating spectra for MEG and HEG, respectively. The result of this script is a single spectrum for MEG and HEG that contains all the observations listed in Table \ref{tab:obs} along with the associated arf, rmf and background files. We calculated the line diagnostics in {\sc xspec} \citep{Arnaud+1996}.
The spectrum were then transformed into a {\sc spex v3.08.01}  \citep{Kaastra+2024} readable format using {\sc trafo v.1.04}\footnote{\url{https://spex-xray.github.io/spex-help/tools/trafo.html}}. 

\section{Line diagnostics}
\label{sec:line}
\begin{figure}
    \includegraphics[width=\linewidth]{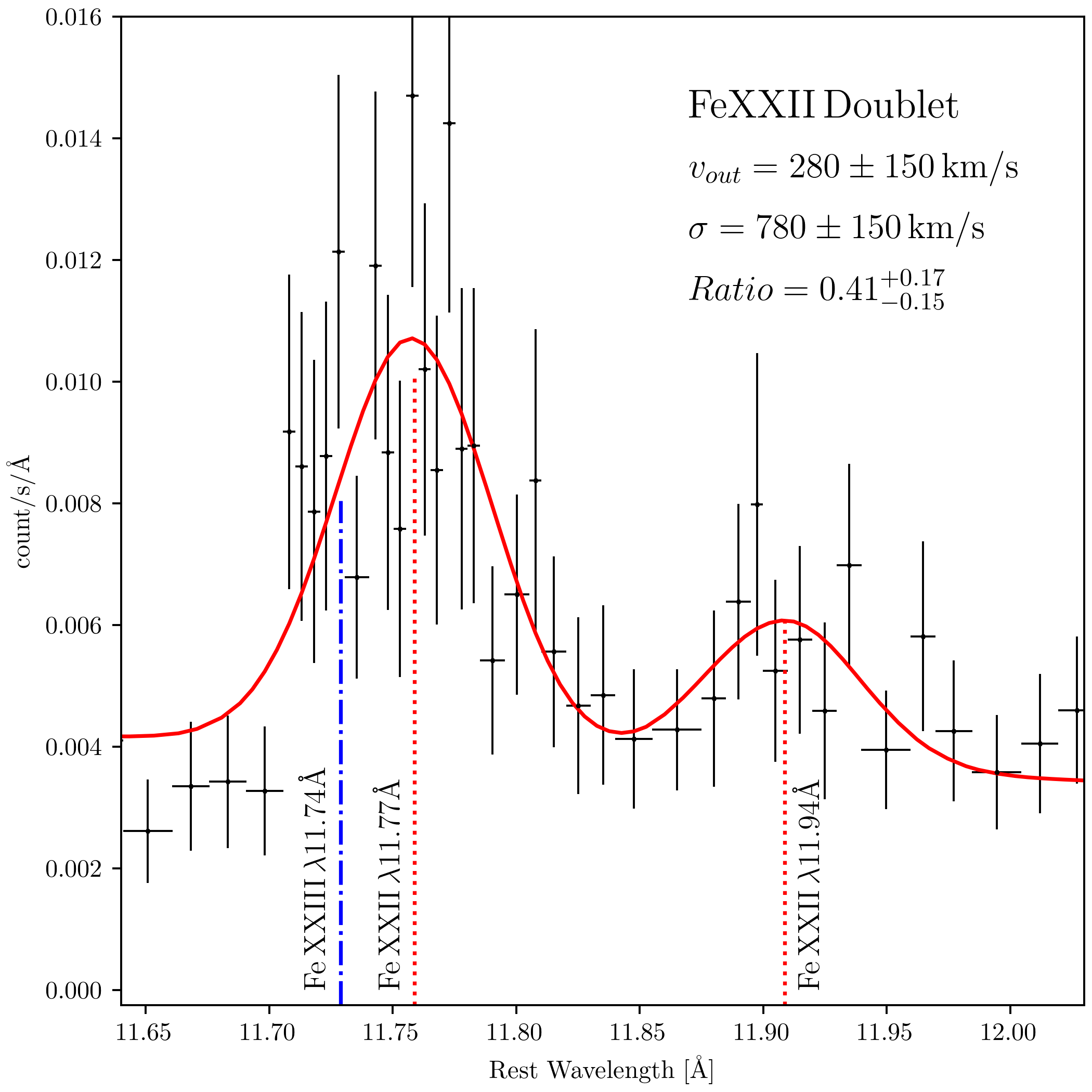}
    \caption
    {The Fe XXII doublet viewed in isolation. The red dotted lines indicate the wavelengths of the \fe\ doublet used by \cite{Mauche+2003, Mauche+2004}. The first blue dot-dashed line ($11.74$\,\AA) is the result of Fe\,{\sc xxiii} emission that can contribute flux to the density diagnostic \cite{Chen+2004, Liang+2008}. The fit shown includes two Gaussian components ({\sc zaguss}) at $11.77,\,11.94$\,\AA. The outflow velocity was $v_{out}=280\pm150$\,km/s, and the ratio of the line strength was $Ratio = 0.41^{+0.17}_{-0.15}$. The width of each line was measured to be $\sigma=780\pm 150\, {\rm km/s}$}
    \label{fig:doublet}
\end{figure}

The diagnostic region of interest for the \fe\ doublet contains multiple unresolvable lines. As was done by \cite{Mauche+2003}, we take the collection of line transitions and label them as the intensity at $11.77\,$\AA\ and $11.92\,$\AA. We begin by examining the line ratio $I$(11.92\,\AA)/$I$(11.77\,\AA). This was done by fitting a very narrow region of the data, between $11.6-12.05$\,\AA, with two Gaussians and a power law. The energy of each Gaussian was fixed to the energy of the \fe\ lines of interest; the width, outflow velocity and normalization were linked. A scaling constant was applied to the 11.92\,\AA\ line and allowed to vary. The results of this fitting procedure are shown in Figure \ref{fig:doublet}, where the data have been binned for clarity. The errors have been calculated within a 90 percent confidence interval. 

The lines are outflowing at $v_{out}=-280\pm150\kms$, with a width of $\sigma=780\pm 150\kms$ and an intensity ratio of $I(11.92\,$\AA$)/I(11.77\,$\AA$)=0.41^{+0.17}_{-0.15}$. The line ratio calculated is consistent with the lower limits of the expected line ratio of $I$(11.92\,\AA)/$I$(11.77\,\AA) \citep{Mauche+2004, Chen+2004, Liang+2008}. This indicates that we can only derive an upper limit for the density of the region emitting the \fe\ lines. This density is $n< 3\times10^{13}\cmc$. We note here again that this limit was calculated for a plasma in collisionally ionized equilibrium, or a blended plasma.

The width of these lines is greater than the turbulent velocity used previously for these data  \cite[600 $\kms$,][]{Kallman+2014}. This could be due to an intrinsic difference in the emitting material, or it could be the manifestation of the blended lines. If we add a Gaussian at 11.74\,\AA\ with the same strength as 11.77\,\AA\ to account for emission from Fe\,{\sc xxiii} \citep{Chen+2004, Liang+2008}, the turbulent velocity becomes $\sigma=630 \pm 180$ and agrees with \cite{Kallman+2014}. More complex modelling is required to further constrain the density of the material responsible for the \fe. 

\section{Spectral Fitting}
\label{sec:spec}
The underlying assumption of the \fe\ diagnostic in the literature is that the material is collisionally ionized. This is a fair assumption for a cataclysmic variable star. However, we wish to explore how these diagnostic lines behave in a plasma photoionized by an AGN. To achieve this, we utilize the photoionization code {\sc pion} \citep{Miller+2015, Mehdipour+2016, DiGesu+2017} and explore a narrow range surrounding the \fe\ doublet, $11-13$\,\AA. While the RGS data for this source has been fit with {\sc pion} \citep{Grafton+2021}, we note that these HETG data have not. 

{\sc pion} requires an input SED to calculate the photoionized emission spectra. We build an SED for the central engine of \ngc\ to model the ionizing continuum, using  {\sc xspec}. We implemented it in {\sc spex} as a file model. It comprised two components: an accretion disk and an X-ray corona. 
The file model was scaled such that the ionizing luminosity between $1-1000$\,Ry ($0.0136-13.6$ keV, \citealt{Tarter+1969}) was $L_{ion} = 10^{44}\,\ergps$. This is the same ionizing luminosity used by \cite{Kallman+2014}. \cite{Grafton+2021} used a similar value of $L_{ion}=1.5\times10^{44}\,\ergps$ in their analysis of the RGS spectrum.

For the accretion disc, we used {\sc NTsed}\footnote{\url{https://github.com/adamgonzalez/NTsed}} (Gonzalez, in prep.), which models the UV/soft X-ray spectrum originating from an accretion disk, as described by \cite{Novikov+1973}. This is the same disc as used by {\sc AGNSED} \citep{Kubota+2018} and the input parameters are black hole mass (M$_{\rm SMBH}$), Eddington accretion rate (${\rm \dot{M}_{Edd}}$), black hole spin ($a$), inner and outer disc radius ($r_{in}$, $r_{out}$), inclination, colour temperature correction factor ($f_{col}$), and co-moving distance to the source ($D$). We assumed M$_{\rm SMBH}=10^7\Msun$ \citep{Greenhill+1997, Lodato+2003, Gallimore+2023}, ${\rm \dot{M}_{Edd}}=0.1$, $a=0.998$, $r_{in}$ to be at the innermost stable circular orbit ($r_{in}=r_{isco}=1.2r_g$; where $r_g$ is the gravitational radius\footnote{The gravitational radius is defined as $r_g= GM/c^2$, where $G$ is the Gravitational constant, $M$ is the mass of the object and $c$ is the speed of light.}) outer disc radius fixed at $r_{out}=10^5\,r_g$, a colour temperature correction of $f_{col}=1.7$ \citep{Done+2012} and $D=14.4\, {\rm Mpc}$ \citep{Bland-Hawthorn+1997}. We assume that the emitting material is above the accretion disc, and sees the disc at a low inclination value ($5^{\rm o}$). This results in a  $0.2-2$ keV luminosity of the disc-black-body component is $L_{0.01-2\, {\rm keV}}=3.1\times10^{43}\, {\rm erg\,s}^{-1}$. 

\begin{figure}
    \includegraphics[width=\linewidth]{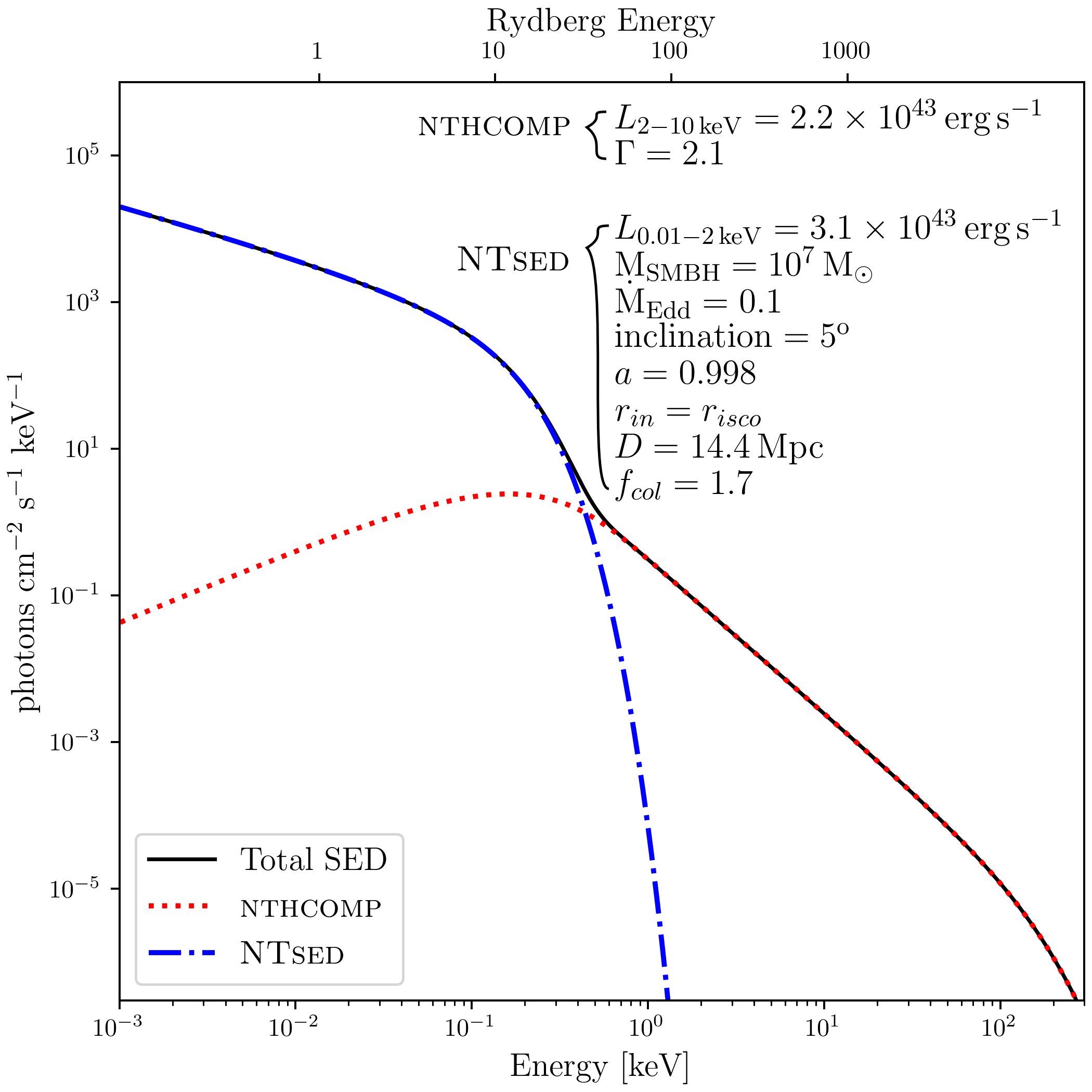}
    \caption
    {The SED used for {\sc pion}  in  this work. The construction of the SED is described in the text.}
    \label{fig:SED}
\end{figure}

The power law component of the SED is modelled using {\sc nthcomp} for a thermally Comptonized continuum. The photon index and $2-10$ keV continuum luminosity were taken from measurements of \ngc\ made with \textit{NuSTAR} \citep{Bauer+2015} so we adopt $\Gamma=2.1$ and $L_{2-10\, {\rm keV}}=2.2\times10^{43}\,\ergps$. The seed photon temperature is fixed at 0.1 keV, and the electron temperature is fixed at 100 keV.  The optical-to-X-ray slope ($\alpha_{\rm OX}$) for this SED was calculated to be $\alpha_{\rm OX}=-0.9$ \citep[e.g.][]{Lusso+2010}.
Our SED is defined between $1\,{\rm eV}-300\,{\rm keV}$ and is shown in Figure \ref{fig:SED}. The values discussed are in general agreement with the review of \ngc\ by \citep{Padovani+2024}. 

We fit the data using {\sc spex v3.08.01} \citep{Kaastra+2024}. The data were optimally binned in {\sc spex} \citep{Kaastra+2016}, and the fit statistics were measured using C-stat \citep{Kaastra+2017}. The Bayesian Information Criterion   \citep[BIC,][]{Schwarz+1978} will be used to compare models. It is given by:  

\begin{equation}
    \textrm{BIC}= k{\rm ln}( n ) + \textrm{C-stat} \, .
    \label{eq:BIC}
\end{equation} 
Here, $k$ is the number of fitted parameters, and $n$ is the number of data points fitted in the spectra. We use $6\leq \Delta\textrm{BIC} \leq10$ to indicate strong evidence and $\Delta\textrm{BIC} \geq10$ to indicate decisive evidence one model is favoured over another \cite[see][]{Kass+1995, Pouliasis+2020}.

Following the iterative method of \cite{Grafton+2021}, the data are first fit with a fiducial continuum model. A photoionized emission component ({\sc pion}) is then added, and a new best fit is found. The parameters are frozen, and another pion component is added. This continues until there is no fit improvement for an additional {\sc pion} component. We do this in an effort not to over-fit the data with endless additions of {\sc pion}. 

The powerlaw continuum is here to describe the general shape of all the weak and unresolved lines originating from plasma components that we do not directly attempt to model. It is true that there are degeneracies between $\Gamma$ and other model parameters, such as powerlaw normalization. Fixing the photon index to $\Gamma=1$ will reduce the statistical errors when calculating uncertainties; the effect is minor for all parameters except the powerlaw normalization. However, this condition is too strict during initial testing, and when exploring the $\Omega-\nh$ degeneracy (Section \ref{sec:discussion}). Thus, for the sake of consistency, and the lack of a physically justifiable reason, we leave $\Gamma$ free.

When initially fitting with {\sc pion}, we keep the absorbing covering fraction ($F_{cov}$) fixed at zero, density fixed at $n=10^4\,\cmc$ matching \cite{Kallman+2014}. We leave the emitting covering fraction ($CF = \Omega/4\pi$) fixed at $\Omega=1$ due to the degeneracies between $\Omega$ and column density \citep[e.g.][]{DiGesu+2017}.  The validity of this is explored later in Section \ref{sec:discussion}. We keep the turbulent velocity of all components fixed to $v_{turb}=600\,\kms$. For each of the {\sc pion} components, we left the column density ($\nh$), outflowing velocity ($v$), and ionization parameter (log\,$\xi$) free to vary.

Fitting the data with a powerlaw (two fit parameters) gives ${\rm C}=2096$ for 197 degrees of freedom (${\rm dof}$), ${\rm BIC}=2106.8$
Adding one {\sc pion} component improves the fit   ($\cdof=768/194$, ${\rm BIC}=794.1$)  significantly  ($\Delta{\rm BIC}=1312.7$). The ionization of this component is ${\rm log}\,(\xi/\ergpcms)=2.3$. This turns out to be the lowest ionization component we find; thus, it is named Low. We now freeze the parameters of the Low ionization component and add a second {\sc pion} component. The second {\sc pion} reduces the ${\rm BIC}=580.4$, again improving the fit significantly ($\Delta{\rm BIC}=268.7$). The ionization of this component is ${\rm log}\,(\xi/\ergpcms)=3.7$, and this is the middle ionization (Mid) component. 

We then added a third {\sc pion} component, and again, it improved the fit significantly (${\rm BIC} = 558.3$; $\Delta{\rm BIC}=17.7$). This component had the highest ionization, ${\rm log}\,(\xi/\ergpcms)=3.9$, and is named High. We tested the need for a fourth {\sc pion} component. However, this had no improvement in the C-stat for three additional fit parameters. Thus, it produced a significantly worse fit (${\rm BIC} =574.2;\,\Delta{\rm BIC}=-15.9$, where the negative indicates the previous model was superior).

Allowing parameters for all three {\sc pion} components free to vary produced the best fit thus far with  $\cdof=419/188$ (${\rm BIC}=477.5$). The main physical difference from the previous model is that the ionization of all three components is lower ($\Delta {\rm BIC}=80.8$). 

\begin{figure}
    \includegraphics[width=1\linewidth]{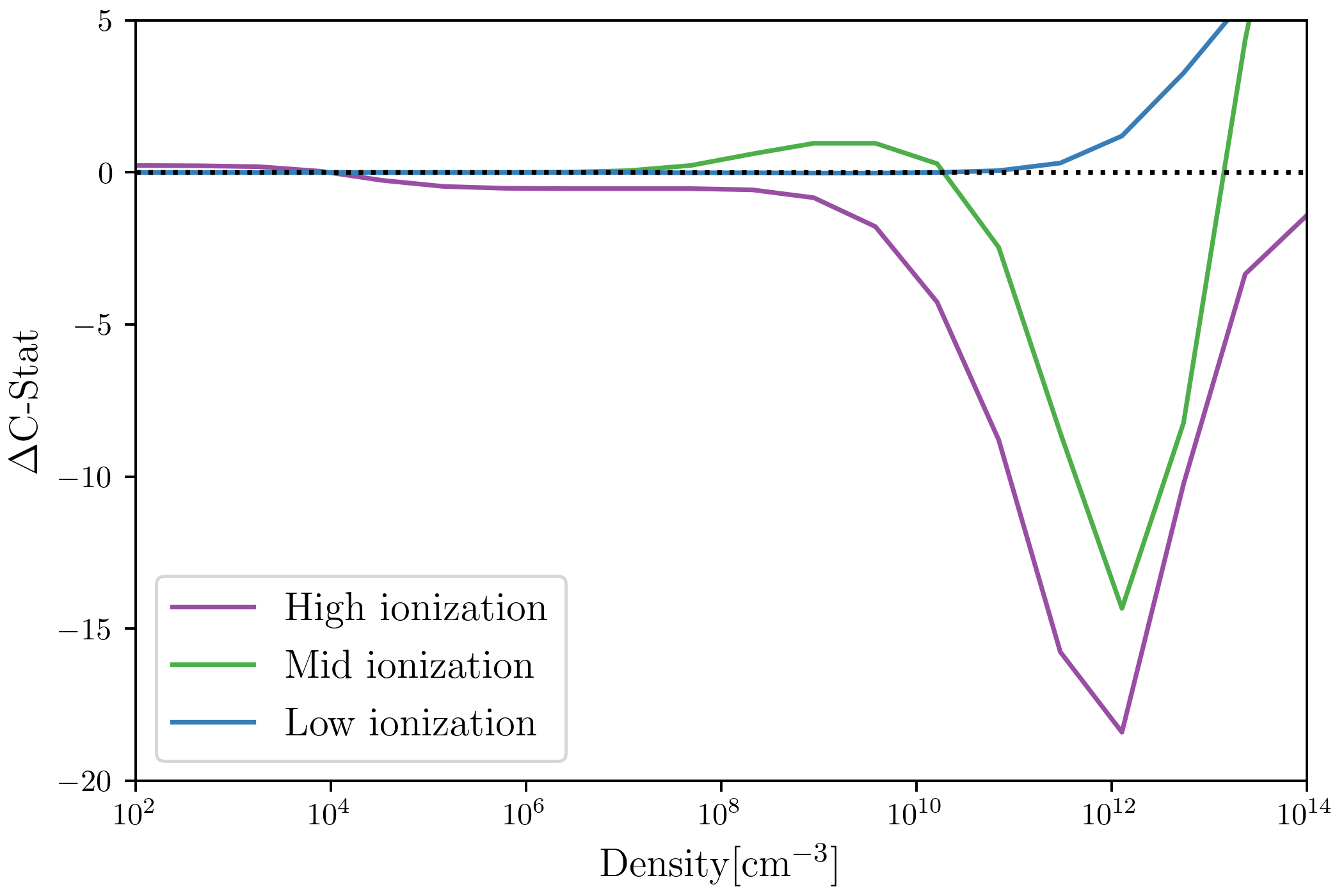}
    \caption
    {Change in fit statistic versus density for each photoionized component. The C-stat is measured relative to the best-fit pion model.  }
    \label{fig:CvD}
\end{figure}

Next, we probe the density of each photoionized component to compare with the upper limit determined by the \fe\ doublet line ratio (Section \ref{sec:line}). We test the density parameter space by stepping through the density of each ionization component from $n=10^2\rightarrow 10^{14}\,\cmc$ with 20 logarithmic steps. The results of this search in parameter space can be seen in Figure \ref{fig:CvD}.

As the density increases, the fit statistic improves for the High ionization component and, to a lesser extent, for the Mid ionization component. The opposite is true for the Low-ionization components, as the fit statistic degrades with increasing density. A density $n< 10^{10}\,\cmc$ is preferred for the Low-ionization component.
The highest densities $> 10^{13}\,\cmc$ are disfavoured by all three components, but especially for the Low- and Mid ionization components. This agrees with our line diagnostic analysis, which excluded densities above $3 \times10^{13}\,\cmc$.

Densities around $\sim10^{12}\,\cmc$ are preferred for the High ionization component. This also appears to be the case for the Mid ionization components. However, this might be an artifact of the complex model fitting when varying only the density of one component at a time and fixing the others. That is, when the High ionization component is fixed at a low density, the Mid ionization component finds a local minimum at higher densities. However, when both components are free to vary, the density of the Mid ionization component becomes very small while that of the High ionization component remains high.

\begin{figure*}
    \includegraphics[width=\linewidth]{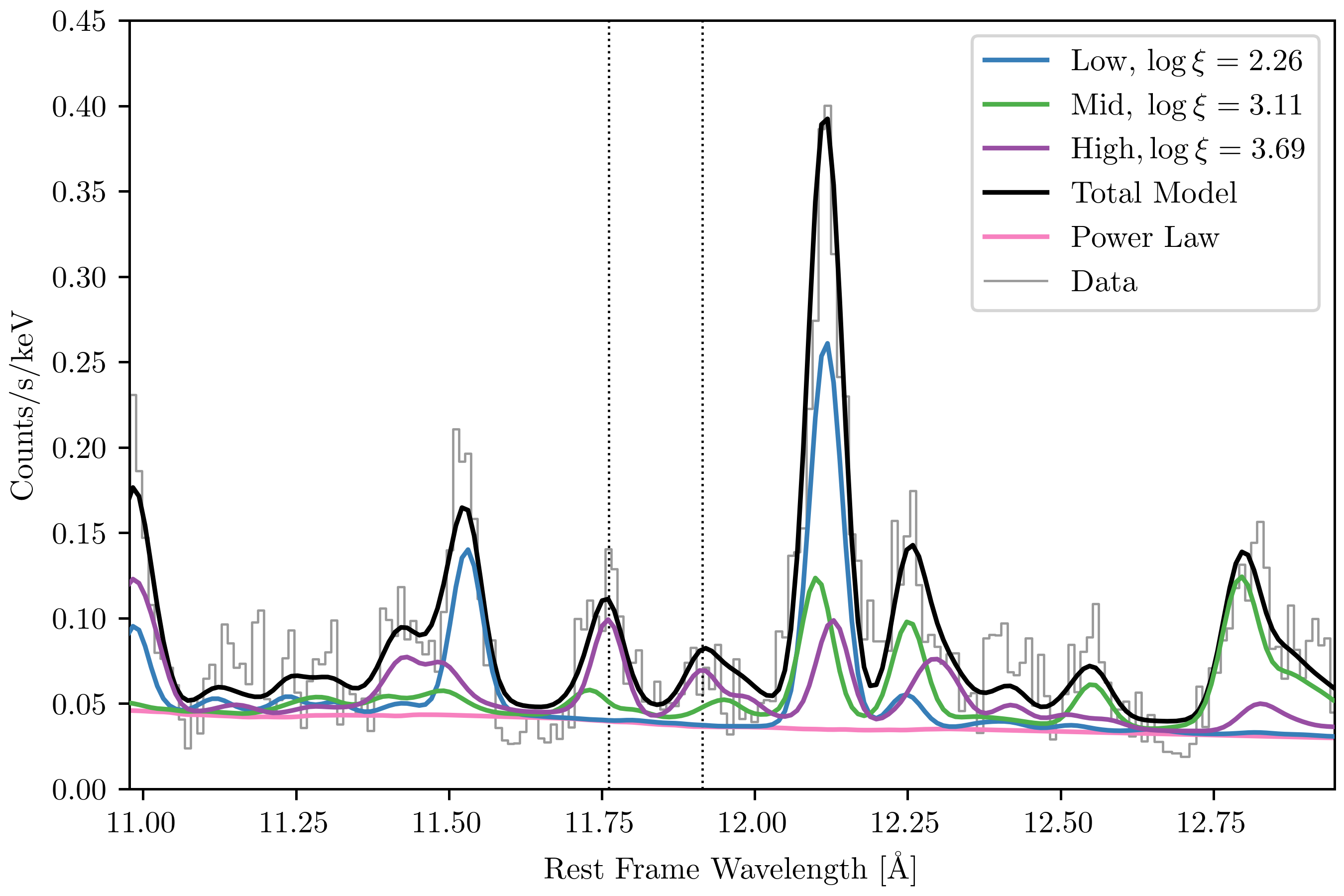}
    \caption
    {Best-fit {\sc pion} model for \ngc. The data is shown in grey and presented without error bars for clarity.  Note that the Fe\,{\sc xxii} doublet is primarily fit by the High ionization components with limited contributions from the Mid ionization components. The black dotted vertical lines denote the location of the 11.77\,\AA\ and 11.92\,\AA\ lines. They have been corrected for the outflow velocity of the High ionization component.} 
    \label{fig:fits}
\end{figure*}

\begin{table}
\centering
\begin{tabular}{c c c c   }
\hline
      (1)       &    (2)     &   (3)             &  (4)  \\
Comp.           & Para.      &  Units            & Value \\ 
\hline
High Ion.       & log$\xi$   & [erg cm s$^{-1}$] & $3.69\pm 0.08$   \\ 
                & $v$        & [$\kms$]          & $-160_{-140}^{+150}$\\
                & $n$        & [$\cmc$]          & $(8.7_{-5.5}^{+12.8})\times 10^{11}$\\
                & $\nh$      & [$\cms$]          & $(1.1_{-0.3}^{+0.5})\times10^{21}$\\
\cline{2-4}
             &$L_{em}$ & [$\ergps$]        & $8.6\times10^{40}$ \\  
                & $\Lambda$  & [erg\,cm$^3$\,s$^{-1}$]&$9.1\times10^{-24}$\\
                & $EM$       & [$\cmc$]          & $9.5\times10^{63}$ \\
\hline
Mid Ion.        & log$\xi$   & [erg cm s$^{-1}$] & $3.11\pm 0.07$  \\ 
                & $v$        & [$\kms$]          & $-820_{-170}^{+160}$\\
                & $\nh$      & [$\cms$]          & $(3.2\pm+0.6)\times10^{20}$ \\
\cline{2-4}
          & $L_{em}$   & [$\ergps$]        & $5.9\times10^{40}$ \\
                & $\Lambda$  & [erg\,cm$^3$\,s$^{-1}$]&$7.9\times10^{-24}$\\
                & $EM$       & [$\cmc$]          & $7.5\times10^{63}$ \\     
\hline
Low Ion.        & log$\xi$   & [erg cm s$^{-1}$] & $2.26\pm 0.07$  \\ 
                & $v$        & [$\kms$]          & $-420\pm 90$\\
                & $\nh$      & [$\cms$]          & $(3.5_{-0.4}^{+0.5})\times10^{20}$ \\
\cline{2-4}              
             &$L_{em}$ & [$\ergps$]        & $4.2\times10^{41}$ \\
                & $\Lambda$  &[erg\,cm$^3$\,s$^{-1}$] &$1.3\times10^{-23}$ \\
                & $EM$       & [$\cmc$]          & $3.2\times10^{64}$ \\
\hline
power law       & $\Gamma$   &                   & $1.3^{+1.4}_{-1.6}$\\
                & norm.      & [ph s$^{-1}$ keV] & $( 3.8\pm0.5)\times10^{49}$ \\

\hline	
\end{tabular}
\caption{Best fit parameters for the emission model of \ngc. Column (1) describes the component name, and columns (2) and (3) describe the fit parameter and the associated units, respectively. Column (4) states the value of the parameters. The luminosity, cooling rate coefficient and $EM$ of each component have also been included.  }
\label{tab:fits}
\end{table}
 
Thus, we leave the density of the highest ionization component free and find a new best-fit of $\cdof=401/187$ (${\rm BIC}=465.0$, $\Delta {\rm BIC}=13.2$). The density of the Low and Mid ionization components remained fixed at $n= 10^4\,\cmc$. The best-fit model is shown in Figure \ref{fig:fits}, and the parameters of this best-fit model are shown in Table \ref{tab:fits}. We note that the density of the High ionization component can be constrained. We can see in Figure \ref{fig:fits} that the High ionization component contributes the most emission to the \fe\ doublet, with some contribution from the Mid ionization component.  We find that $\sim30\%$ of the total line flux between $11.65-12.00\,$\AA\ originates from the Mid component. Less than $\sim 20\%$ of the line flux at $11.77\,$\AA\ originates from the Mid component.   

We find $\Lambda$, the cooling rate coefficient \citep[e.g.][]{Kallman+1982} for each component using the {\sc spex} subroutine {\sc heat} and the density of the plasma. The emitting luminosity ($L_{em}$) is calculated by {\sc spex} between $0.0136-13.6$ keV. $\Lambda$ and $L_{em}$ were then used to calculate the emission measure ($EM$) of each component using Equation \ref{eq:EM} \citep[see][Eq. 4]{Fabian+1994}:
\begin{equation}
    EM  = \frac{L_{em}}{\Lambda}
    \label{eq:EM}
\end{equation}
The emission measure, cooling rate coefficient and emitting luminosity of each {\sc pion} component are in Table \ref{tab:fits}. The cooling rates provided by {\sc spex} are in good agreement with those found in \citep{Kallman+1982}.  The $EM$ of these components is also in good agreement with those found by \cite{Kallman+2014} and \cite{Grafton+2023} in the HETG and RGS data, respectively. We note, however, that the emission measure calculated using Eq. 1 from \cite{Grafton+2021} produced a value more than two orders of magnitude smaller than those we report in Table \ref{tab:fits}. If this $EM$ were trusted, then the cooling rate required based on the emission luminosity would be significantly higher than what is expected based on the literature and the calculation from {\sc spex}. We suspect this discrepancy in $EM$ is due, in part, to the small column density and high $\Omega$. When the data is fit with column density fixed at $\nh=5\times10^{23}\,\cms$ and $\Omega$ free (see Section \ref{sec:discussion} for more details), the emission measure is still an order of magnitude smaller than literature values. Thus, perhaps using the cooling rate and emission luminosity is a more accurate method for calculating emission measure.  It utilizes information from the model as a whole instead of select model parameters (e.g. $\xi$, $\Omega$, $\nh$).

\section{Fe {\sc xxii} Diagnostic}
\label{sec:diagnostic}
\begin{figure*}[ht!]
    \includegraphics[width=1\linewidth]{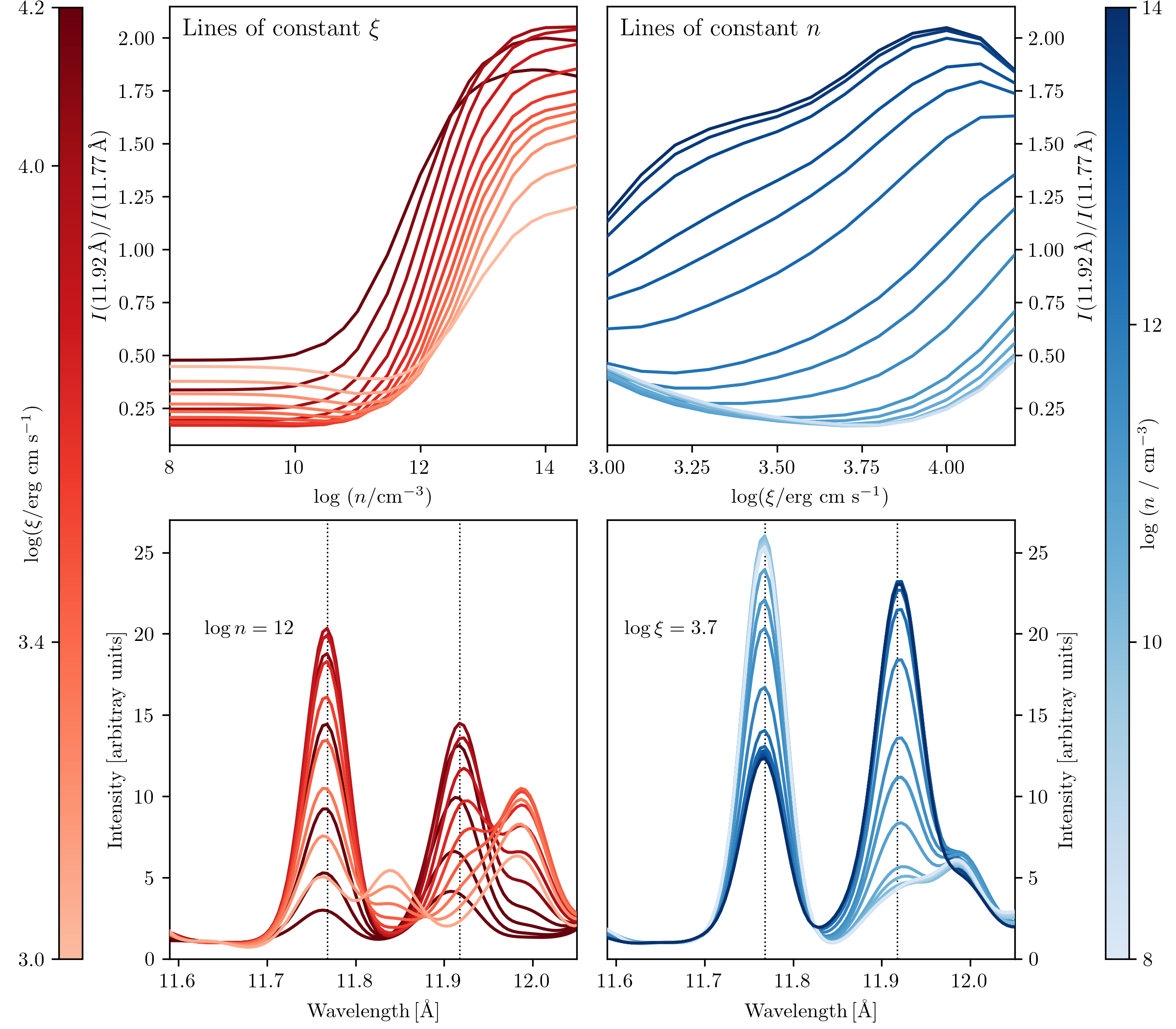}
    \caption
    {Fe\,{\sc xxii} diagnostic curves and line profiles generated using {\sc pion} for different ionization (log\,$\xi$) and densities ( $n$). The SED used is representative of NGC\,1068. \textit{Top left panel:} Density verse the ratio of line strengths,  $I(11.92$\,\AA$)/I(11.77$\,\AA$)$ at different ionizations. Higher values of log\,$\xi$ are shown in darker red. \textit{Bottom left panel:} the line profile of the Fe\,{\sc xxii} doublet at different values of log $\xi$. Density is fixed at $n = 10^{12}\, {\rm cm}^{-3}$. The dotted black lines show the location of each of the diagnostic lines. Note that at this density, the presence of the 11.92\,\AA\ emission is ionization dependant. The 11.92\,\AA\ emission is present at lower ionization when the density is higher (not shown here). \textit{Right top panel:} Ionization verse $I(11.92$\,\AA$)/I(11.77$\,\AA$)$ at different densities. Higher values of $n$ are shown in darker blue. \textit{Bottom right panel:} the Fe\,{\sc xxii} doublet at different values of $n$ with ionization fixed at log\,$(\xi/\ergpcms) = 3.7$. At lower densities, the $11.77\,$\AA\ line is very strong, and the $11.92\,$\AA\ line is relatively weak. At higher densities, the reverse is true.}
    \label{fig:bigFig}
\end{figure*}
\begin{table*}
\centering
\begin{tabular}{l r l c c }
\hline
      (1)     &  \multicolumn{2}{c}{(2)}      &   (3)       &(4)\\
Ion           & \multicolumn{2}{c}{Electron Transitions}     &$\lambda$    & Relative \\ 
              &                               &    &    \AA  & Intensity  \\ 
\hline
Fe {\sc xxii}  & $2s^1 2p^1 (^3{\rm P}) 3p^1 \,{ }^2{\rm P}_{3/2}\,\,\, \rightarrow$ & $2s^2 2p^1 \, { }^2{\rm P}_{3/2}$ & 11.6549 & 0.02 \\ 
Fe {\sc xxii}  & $2s^1 2p^1 (^3{\rm P}) 3p^1 \,{ }^4{\rm D}_{3/2}\,\,\, \rightarrow$ & $2s^2 2p^1 \, { }^2{\rm P}_{3/2}$ & 11.7051 & 0.02 \\ 
Fe {\sc xxii}\, ($\dagger$) & $2s^2 3d^1 \,{ }^2{\rm D}_{3/2}\,\,\,     \rightarrow$ & $2s^2 2p^1 \, { }^2{\rm P}_{1/2}$ & 11.7674 & 1.00 \\ 
Fe {\sc xxiii} & $2s^1 3s^1 \,{ }^3{\rm S}_1\,\,\,                      \rightarrow$ & $2s^1 2p^1 \, { }^3{\rm P}_2$     & 11.8733 & 0.02 \\ 
Fe {\sc xxiii} & $2s^2 2p^1 3d^1 \,{ }^1{\rm P}_1\,\,\,                 \rightarrow$ & $2s^2 2p^2 \, { }^1{\rm S}_0$     & 11.8970 & 0.08 \\ 
Fe {\sc xxii}\, ($\dagger$) & $2s^2 3d^1 \,{ }^2{\rm D}_{5/2}\,\,\,     \rightarrow$ & $2s^2 2p^1 \, { }^2{\rm P}_{3/2}$ & 11.9208 & 0.48 \\ 
Fe {\sc xxii}  & $2s^2 3d^1 \,{ }^2{\rm D}_{3/2} \,\,\,                 \rightarrow$ & $2s^2 2p^1 \, { }^2{\rm P}_{3/2}$ & 11.9335 & 0.05 \\ 
Fe {\sc xxi}   & $2s^1 2p^2 (^4{\rm P}) 3p^1 \,{ }^3{\rm D}_1\,\,\,     \rightarrow$ & $2s^2 2p^2 \, { }^3{\rm P}_0$     & 11.9464 & 0.02 \\ 
Fe {\sc xxi}   & $2s^1 2p^2 (^4{\rm P}) 3p^1 \,{ }^3{\rm P}_2\,\,\,     \rightarrow$ & $2s^2 2p^2 \, { }^3{\rm P}_2$     & 11.9793 & 0.05 \\ 
Fe {\sc xxi}   & $2s^1 2p^2 (^4{\rm P}) 3p^1 \,{ }^3{\rm P}_1\,\,\,     \rightarrow$ & $2s^2 2p^2 \, { }^3{\rm P}_2$     & 11.9860 & 0.03 \\ 
Fe {\sc xxi}   & $2s^1 2p^2 (^4{\rm P}) 3p^1 \,{ }^5{\rm P}_1\,\,\,     \rightarrow$ & $2s^2 2p^2 \, { }^3{\rm P}_0$     & 11.9906 & 0.07 \\ 
Fe {\sc xxi}   & $2s^1 2p^2 (^4{\rm P}) 3p^1 \,{ }^3{\rm D}_3\,\,\,     \rightarrow$ & $2s^2 2p^2 \, { }^3{\rm P}_2$     & 11.9924 & 0.08 \\ 
Fe {\sc xxi}   & $2s^1 2p^2 (^4{\rm P}) 3p^1 \,{ }^3{\rm D}_2\,\,\,     \rightarrow$ & $2s^2 2p^2 \, { }^3{\rm P}_1$     & 12.0086 & 0.07 \\ 
Fe {\sc xxi}   & $2s^1 2p^2 (^4{\rm P}) 3p^1 \,{ }^3{\rm D}_1\,\,\,     \rightarrow$ & $2s^2 2p^2 \, { }^3{\rm P}_1$     & 12.0528 & 0.03 \\ 
\hline	
\end{tabular}
\caption
{Emission lines relevant to the \fe\ density diagnostic. Column (1) the iron ion, Column (2) the electron configuration before and after the transition. Column (3) the associated wavelength used by {\sc spex}. Column (4) gives the relative strengths of each emission line compared to the strength of the $11.77\,$\AA\ line. These values are calculated for the best-fit scenario only, as they change with ionization and density. ($\dagger$) These are the two line transitions that are commonly identified as driving the \fe\ line diagnostic. Here we list the wavelength of each line to 4 digits after the decimal, as used by {\sc pion}. However, this precision far exceeds the energy resolution of HETG. In the text, we always state the wavelength to three digits after the decimal. }
\label{tab:iron}
\end{table*}

The density of the High ionization plasma appears to be very well constrained using the spectral model. This is at odds with the predictions made using the line ratio and the literature values of the \fe\ diagnostic (see Section \ref{sec:line}). To explore further, we generated diagnostic curves similar to those produced in the literature \citep[e.g.][]{Mauche+2004} using {\sc pion}. This was achieved by calculating a model spectrum of a single {\sc pion} component at various densities and ionization states. We then calculated the $I(11.92$\,\AA$)/I(11.77$\,\AA$)$ line ratio under these various conditions using the model spectra. Table \ref{tab:iron} lists all the prominent (greater than 2\% the strength of $I$(11.77\,\AA)) emission lines between $\sim11.65-12.05\,$\AA\ used by {\sc pion} to generate the simulated spectra for the best-fit parameters for density and ionization. These lines contribute to the $I(11.92$\,\AA$)/I(11.77$\,\AA$)$ line ratio, and are formed by transitions of Fe\,{\sc xxi}\,--\,Fe\,{\sc xxiii}. The exact strength of each line for every combination of density and ionization is different. Table \ref{tab:iron} is included to illustrate the complexity of the iron transitions in this very narrow wavelength region.

We found that the strength of the diagnostic ratio depended primarily on the ionization state and density of the emitting regions. The top left panel of Figure \ref{fig:bigFig} shows the diagnostic curves similar to those seen in \cite{Mauche+2003, Mauche+2004}. However, the critical density (where the diagnostic is most sensitive to density) is lower and decreases with increasing ionization. Thus, we conclude that using an AGN SED  (power law + disc component) instead of a collisionally ionized plasma lowers the critical density by up to three orders of magnitude, depending on the ionization of the plasma. Furthermore, the lowest limit of the $I(11.92$\,\AA$)/I(11.77$\,\AA$)$ ratio is as low as $\sim0.17$ versus the literature value of $\sim0.4$ \citep{Mauche+2004, Chen+2004}. We attribute these lower values to the ability of the AGN continuum to indirectly but effectively populate the first excited state ($2s^22p^1\, ^2{\rm P}_{3/2}$). Excitation from the first excited state to $2s^2 3d^1 \,{ }^2{\rm D}_{5/2}$ is the primary method to populate this level and produces the 11.92\,\AA\ emission line. In a collisional plasma, the first excited state is populated by collisional processes. For the photoionized plasmas we explore in this work, the first excited state is populated entirely by radiative processes in the low-density limit. As the density of the plasma increases, collisional excitations are necessary for the population of the first excited state to surpass the ground state, thus making the 11.92\,\AA\ emission line strong. The combination of collisional and radiative processes populating the first excited state results in a lower critical density in the presence of an AGN radiation field. We discuss this in more detail in Appendix \ref{apen}. 

Unfortunately, the trends in the diagnostic ratio do not behave monotonically. We can see in the top left panel of Figure \ref{fig:bigFig} at low densities. As ionization increases from low values (light red) to high values (dark red), the ratio decreases in value before increasing again. This ionization dependence can also be seen in the top right panel, where $I(11.92$\,\AA$)/I(11.77$\,\AA$)$ versus log\,$\xi$ has been plotted. The lighter blue lines indicate lower densities, while the darker blue lines represent higher densities.  

By looking at the line profiles in the bottom panels of Figure \ref{fig:bigFig}, we can understand the cause of this behaviour. For comparison, each model spectrum has been normalized to the intensity at $11.65\,$\AA. This was chosen because there are no strong emission lines here. The $11.77\,$\AA\ emission is present in all line profiles. However, the emission at 11.92\,\AA\  is produced only at sufficient ionization and densities. The bottom right panel shows that several Low ionization model spectra have no or limited contributions from 11.92\,\AA. At higher densities, the 11.92\,\AA\ line is present at the lower ionization values. We can also see that at the highest ionization values, the strength of the emission at  11.77\,\AA\ and 11.92\,\AA\ has significantly diminished. This is most likely due to the photoionized plasma becoming over-ionized. 

The bottom right panel of Figure \ref{fig:bigFig} shows the line profile of the doublet with changing density. We see the expected behaviour here, where at low densities (light blue), the emission at 11.77\,\AA\ is strong while emission at 11.92\,\AA\ is weak. As density increases, the 11.77\,\AA\ emission reduces in strength and the emission at 11.92\,\AA\ increases. 

The effects of an AGN SED on the critical density have been indirectly seen before in absorption by \cite{Mao+2017} and  \cite{Tomaru+2023}. The diagnostic curves developed in absorption are most sensitive to densities $\sim10^{12}\,\cmc$, similar to what we find in emission.  

\begin{figure}
    \includegraphics[width=1\linewidth]{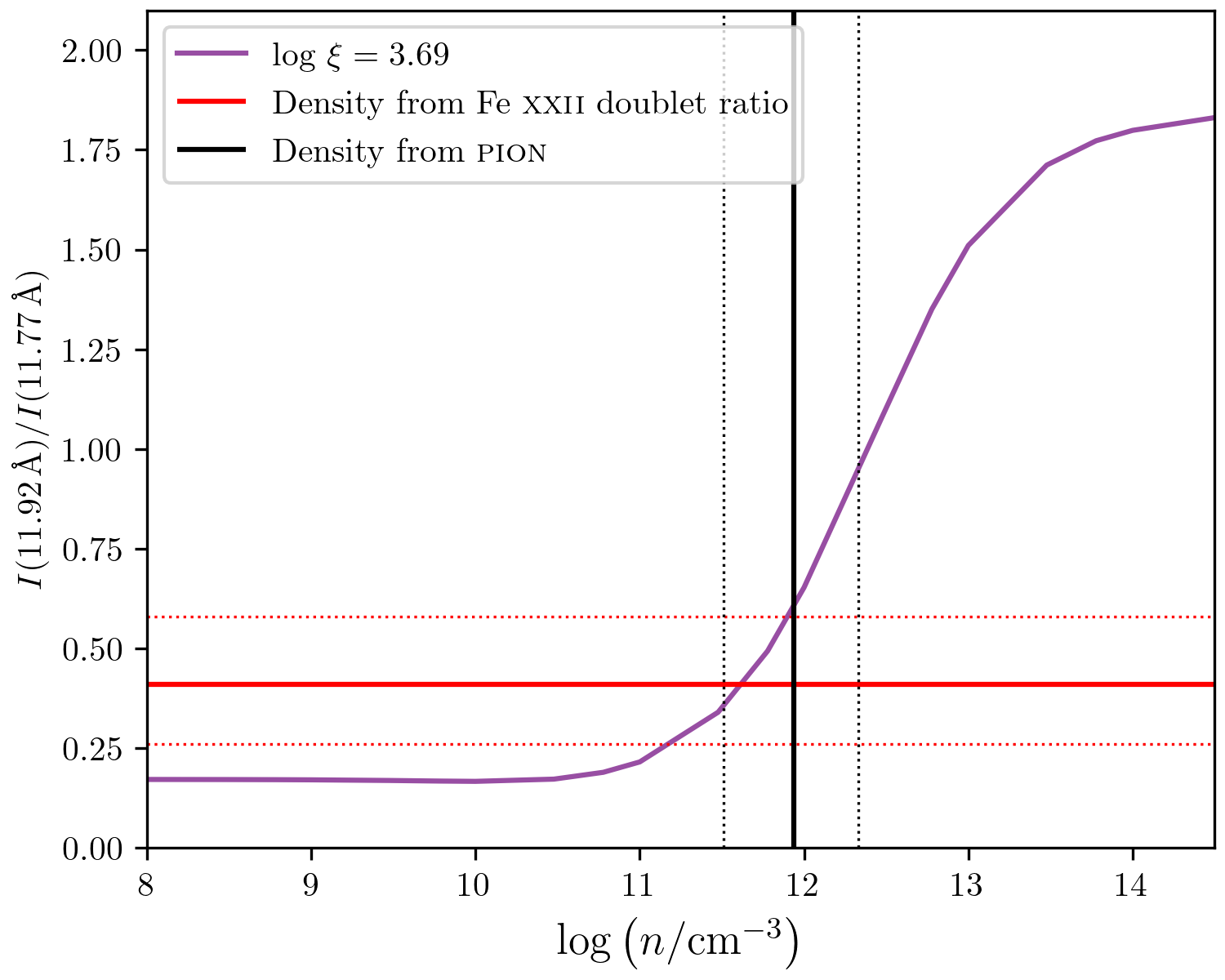}
    \caption
    {Density verse $I(11.92$\,\AA$)/I(11.77$\,\AA$)$ for best-fit {\sc pion} component,  log\,$(\xi/\ergpcms)=3.69$. The red horizontal line shows the measured ratio of the \fe\ doublet from the data. The horizontal dotted lines are the 90\% confidence interval on the measured ratio.  The black vertical line shows the measured density from the High ionization {\sc pion} component.}
    \label{fig:hiXi}
\end{figure}

To demonstrate that the \fe\ doublet is the driver of the density constraint with the \ngc\ data, we generate a diagnostic line with log\,$(\xi/\ergpcms)=3.69$, the best-fit value for the High ionization component. Figure \ref{fig:hiXi} shows this diagnostic ratio with the density constraints placed by both {\sc pion} and doublet indicated by the vertical and horizontal lines, respectively. We can see that the densities predicted by the diagnostic and measured by the model agree within error. There is a discrepancy between these two measurements, however.  The model measured a density $\sim0.4$\,dex larger than the ratio. This can be attributed to contributions from the Mid ionization component (discussed in Sec. \ref{sec:spec}). There may also be other lines in the spectrum between $11-13$\,\AA, which are somewhat density-sensitive \citep[e.g. Fe\,{\sc xxiii} and/or Fe\,{\sc xxi},][]{Mao+2017}. Overall, we can conclude that the density of the High ionization plasma is largely driven by the \fe\ doublet. 

We do not explore the effect column density has on the diagnostic ratio in depth. This is due to the degeneracy between $\Omega$ and $\nh$, as well as the added complexity that a third parameter would introduce. We find that, in initial tests, the column density does not have a significant impact on the diagnostic curves until  $\nh > 10^{22}\,\cms$. After this, the diagnostic curves begin to behave similar to those with increasing ionization values. Further investigation would be needed to classify this behaviour; however, that is outside the scope of this work. 

\section{Discussion} 
\label{sec:discussion}
We first wish to identify the emission radius of the ionized material. We do this by utilizing the definition of ionization, 
\begin{equation}
    \xi = \frac{L_{ion}}{n\,r^2}\, ,
\end{equation}
where $r$ is the distance between the ionization source and the emitting material, we can calculate the distance from where the High ionization component is emitted. The density of the High ionization component is $n = 8.7\times 10^{11}\, \cmc$, and the ionization is $\xi = 10^{3.69} \erg\,{\rm cm\,  s}^{-1}$. The ionizing radius is $\sim 1.5\times10^{14}\,{\rm cm}$. Assuming the black hole mass to be $\sim 10^7 \Msun$ \citep{Greenhill+1997, Lodato+2003, Gallimore+2023}, the emitting region is at $\sim100\,r_g$ away from the ionizing source. An emission line at this radius would be expected to have a velocity broadening due to Keplerian rotation of $\sim30000\, \kms$. The observed broadening is $\sim2$\% of this.

In an attempt to quantify uncertainty on this measurement, we assume that the errors are Gaussian in nature. This is a crude assumption as the error bars derived in Table \ref{tab:fits} are often uneven. However, the Gaussian assumption should be sufficient for an order-of-magnitude approximation of uncertainty. We find the uncertainty in this calculation to be of order $\sim80\,r_g$.

To achieve a true statistical error bar for this measurement, the probability distribution function must be known. This is typically achieved with the use of tools such as MCMC \citep[see][and references therein]{Robert+2008} or Nautilus \citep{nautilus}. Unfortunately, these methods are extremely computationally expensive and not built-in methods within the {\sc spex} framework, requiring implementation with {\sc pyspex}\footnote{\url{https://spex-xray.github.io/spex-help/pyspex.html}}. We do not report the uncertainty on any other calculated value, due to the limitations discussed above. This is not expected to significantly impact the major results of this work.

The column density and density of the High ionization component allow us to calculate the size of the emission region. By assuming $d_H = {\rm N_H}/n$, where $d_H$ is the diameter, we find for ${\rm N_H}=1.2 \times 10^{21}\, \cms$, $d_H\sim 10^9\, {\rm cm}$ ($\sim10^{-3}\, r_g$). This is a rather thin shell of material that would surround the central region uniformly. 

If we instead assume that the column density is $N_H=5\times10^{23}\, \cms$, which is comparable to values of column density in the literature \citep{Kallman+2014, Grafton+2021}, the emission covering fraction, $\Omega$, becomes small. This scenario is consistent with a small observing window in clumpy obscuring material. The value of $\Omega$ in this scenario would be $\Omega \simeq 0.03$ and produces a fit statistic of $\cdof=409/187$ (${\rm BIC}=472.8$). This is greater than the thin shell scenario ($N_H$ free and $\Omega$ fixed) with $\Delta {\rm BIC}=7.8$. However, there is insufficient evidence to decisively exclude this scenario. This column density would have a cloud diameter of $\sim0.4\,r_g$.

At low values of $\nh$, the effects on the line ratios are limited. We note again that only a limited study of the effects of $\nh$ on the line ratios has been performed here. However, it is not until $\nh>10^{22}\,\cms$ that changes to the diagnostic ratio are observable. The changes are most substantial at higher densities. Below this threshold, the column density acts like a sort of normalization factor, having much the same effect as $\Omega$. When we attempted to fit the spectra with $\Omega$ and $\nh$ as free parameters, the data preferred $\Omega > 1$, which is extremely challenging to explain physically. Thus, we fixed $\Omega=1$, and accepted the lower as perhaps a product of the limited band pass we are fitting over.

An emission radius of $\sim100\,r_g$ is challenging to explain given the Seyfert 2 nature of this source. This would place the emission region in the accretion disc and would require a direct line of sight to the central engine. This would be very atypical behaviour for a Seyfert 2 as they are thought to be absorbed completely by an obscuring torus \citep{Urry+1995}. We now explore alternative scenarios to reconcile the diagnostic results with the expected behaviour of this object. 

\subsection{Effects of SED}
\label{sec:ngcSED}
\begin{figure*}
    \includegraphics[width=1\linewidth]{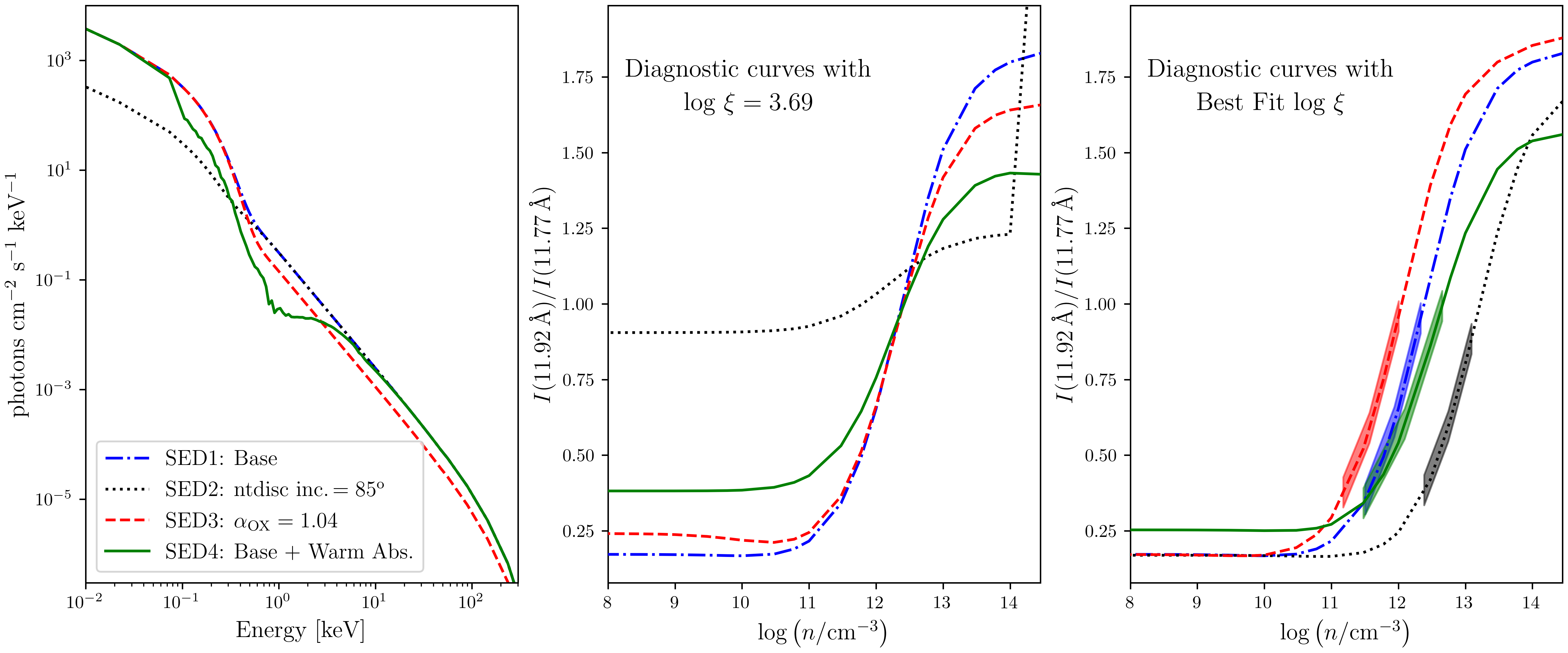}
    \caption
    {\textit{Left panel:} the four ionizing continua used to test the effects of  SED on the \fe\ diagnostic. The construction of each is described in the text. \textit{Middle panel:}  model ratio at 11.77\,\AA\ and 11.92\,\AA\ for each of the SEDs at ${\rm log}\,(\xi/\ergpcms)=3.69$. \textit{Right panel:} diagnostic curves for each SED at the best-fit ionization for each SED. The density of the High ionization component was constrained for each model and is shown in the shaded region.  }
    \label{fig:4xsed}
\end{figure*}

The properties of a photoionized spectrum are deeply dependent on the ionizing continuum that is used \citep{Mehdipour+2015}.  Thus, we explored the effects of four different ionizing continua on the density and ionization measured. We assign the base ionizing continuum constructed in Section \ref{sec:spec} and Figure \ref{fig:SED} as SED1. It is shown in the left panel of Figure \ref{fig:4xsed} with the blue dot-dashed line. The three remaining ionizing continua are variations on SED1. 

The first modification we make reduces the strength and shape of the UV component of the ionizing continuum. This was done by adjusting the inclination of the {\sc NTsed} to $85^{\rm o}$. We label this SED2, which is shown in the left panel of Figure \ref{fig:4xsed} with the black dotted line.  

For SED3, we reduce the strength of the primary X-ray continuum to increase the relative strength of the UV component. The luminosity of the {\sc nthcomp}  component was reduced by $\sim50$ percent to $L_{2-10 \kev}=10^{43}\,\ergps$. The new optical to X-ray slope is  $\alpha_{\rm OX}=-1.04$. SED3 is shown in the left panel of Figure \ref{fig:4xsed} with the red dashed line. 

To simulate the effects of absorption between the central engine and the emitting cloud, SED4 was constructed. It consists of the base ionizing continuum modified by four warm absorbers. The ionization and column density of each absorber were consistent with those found in \ngc\ by \cite{Grafton+2021}; however, the covering fraction of each was increased to $CF=0.8$ to produce a stronger absorption feature. We used the {\sc xspec} model {\sc zxipcf} to apply this absorption to the base model, despite \cite{Grafton+2021} using {\sc pion} components for the emission. This was done because the resultant SED was then binned, washing away any major difference between the two models into a lower-resolution continuum. This SED4 is shown in the left panel of Figure \ref{fig:4xsed} with the green solid line. We note here that SED4 produced the best fit over all, with a $\Delta{\rm BIC}=15.2$. The separate SEDs were applied to all three emission components, so this change in the fit statistic could be due, in part, to improvements in the fit from the other two components.  

The diagnostic curves were calculated using the previous method for a single ionization value of $\xi=10^{3.69} \, \erg\,{\rm cm\,  s}^{-1}$. This is shown in the middle panel of Figure \ref{fig:4xsed}. We see different behaviour for each of the ionizing continua used. First, comparing SED1 with SED3, the primary difference between these ionizing continua is the reduction in hard X-rays, which increases the $\alpha_{\rm OX}$ slope. This results in a very similar diagnostic curve, with the differences in the highest and lowest densities.

The diagnostic curve for SED4 has a shallower slope and smaller range of values for $I$(11.92\,\AA )/$I$(11.77\,\AA),  compared to SED1 and SED3. At the same ionization parameter, differences in the ionizing continuum appear to have a significant effect on the emission at 11.92\,\AA\ and 11.77\,\AA.

SED2 shows very little of the expected behaviour that the other three ionizing continua display. The low-density limit of the diagnostic ratio is very high ($I(11.92$\,\AA$)/I(11.77$\,\AA$)\sim 0.9$), and there appears to be a discontinuity at $n\sim10^{14}\,\cmc$  This is because at the test ionization of $\xi=10^{3.69} \, \erg\,{\rm cm\,  s}^{-1}$, the \fe\ lines are not present. This curve is included for completeness and to highlight some of the extreme differences that can appear as the SED is modified. 

Next, we apply these ionizing continua to best-fit the pion model parameters and re-fit the data.  The best-fit ionization and density for the High ionization component for each ionizing continuum are found in Table \ref{tab:SED}. We recalculate the diagnostic curves for each SED using their best-fit ionization. The results are plotted in the right panel of Figure \ref{fig:4xsed}.

\begin{table*}
    \centering
    \begin{tabular}{cccccc}
    \hline
(1)     & (2)                             & (3)        &(4)              &(5)        & (6)     \\
Label   & Description                     & $\xi$ & $n$             & $r_{ion}$ & BIC     \\
        &                                 & $\erg\,{\rm cm\,  s}^{-1}$           & $10^{11}\cmc$   &$r_g$      &         \\
   \hline
 SED1   & Base                            & $10^{3.69}$ &$8.7_{-5.5}^{+12.8}$   & 104       & 463.9   \\
 SED2   & {\sc NTsed} inc.\,$=85^{\rm o}$& $10^{3.04}$ &$56_{-31}^{+68}$ & 87        & 469.3   \\
 SED3   & $\alpha_{\rm OX}=1.04$          & $10^{4.01}$ &$3.7_{-2.2}^{+6.4}$    & 111       & 459.2   \\
 SED4   & Base  + Warm Abs.               & $10^{3.67}$ &$12.4_{-9.4}^{+32.6}$  & 89        & $448.7^{\dagger}$   \\
    \hline
    \end{tabular}
    \caption
    {Column (1) listed the label for each ionizing continuum tested. Column (2) gives a brief description of the ionizing continuum, while columns (3), (4), (5), and (6) list the ionization, density, calculated radius, and BIC, respectively, for each ionizing continuum. ($\dagger$) SED4 given the best fit overall.} 
    \label{tab:SED}
\end{table*}
Remarkably, all four ionizing continua tested produce similar diagnostic curves at their best-fit ionization, with the critical densities increasing from SED3, SED1, SED4, to  SED2 ($\alpha_{OX}=1.04$, Base, Base  + Warm Abs., {\sc NTsed} inc.\,$=85^{\rm o}$). 
SED1, SED2, and SED3 have near identical ranges of theoretical $I(11.92$\,\AA$)/I(11.77$\,\AA$)$  values. SED4 has a smaller range of theoretical $I(11.92$\,\AA$)/I(11.77$\,\AA$)$ but, overall, retains the same general shape as the other three curves calculated. However, the slope of diagnostic curves produced by SED1, SED2, and SED3 is the same at their best-fit log $\xi$\ values. The slope of the diagnostic curve for SED4 is shallower than the other diagnostic curves. We note that the difference between SED1 and SED4 is greatest at $\sim1\,{\rm keV}$, which is $\sim$ to the location of the \fe\ doublet. We do not speculate further on the exact cause of this change as it is outside the scope of this work.

The density measured by each SED is shown in the shaded region in Figure \ref{fig:4xsed}, and listed in Table \ref{tab:SED}. We see that the range in density is $\sim 10^{11}-10^{13}\, \cmc$. This is much larger than the range in any individual SED and indicates that the measured density is dependent on the input ionizing continuum. We recalculate the ionizing radius, $r_{ion}$ of the emission region for each ionizing continuum, assuming $L_{ion} = 10^{44}\,\ergps$ and find remarkable consistency in the ionizing radius. All four emission regions are $\sim100\,r_g$ from the ionizing source. This could be a result of the density-ionization degeneracy, shown in Section \ref{sec:diagnostic}. Nevertheless, we can conclude that, under all SEDs tested, the data demands that the distance from the ionizing source is small. 

\subsection{Indirect Methods to Probe the Location of the \fe\ Emission}
Here, we probe two methods external to the ionization and density that could provide an indication of the location of the \fe\ emission. The first is an exploration of any variability from the High ionization component. The second is emission in higher energy bands from the High ionization component.

\subsubsection{Variability Analysis}
\label{sec:Varibility}
The central engine of an AGN is often very variable \cite[e.g.][]{Leighly+1999A, Leighly+1999B, Grupe+2001, Parker+2013, Wilkins+2014, Wilkins+2015B, Buhariwalla+2024}; however, the obscuring torus of a Seyfert 2 galaxy blocks this variability from our sight. \cite{Bauer+2015} concluded that there was no detectable broadband variability from the central engine in \ngc. In this section, we attempt to determine if we are able to detect variability in the \fe\ doublet if any were present.

\begin{figure}
    \includegraphics[width=1\linewidth]{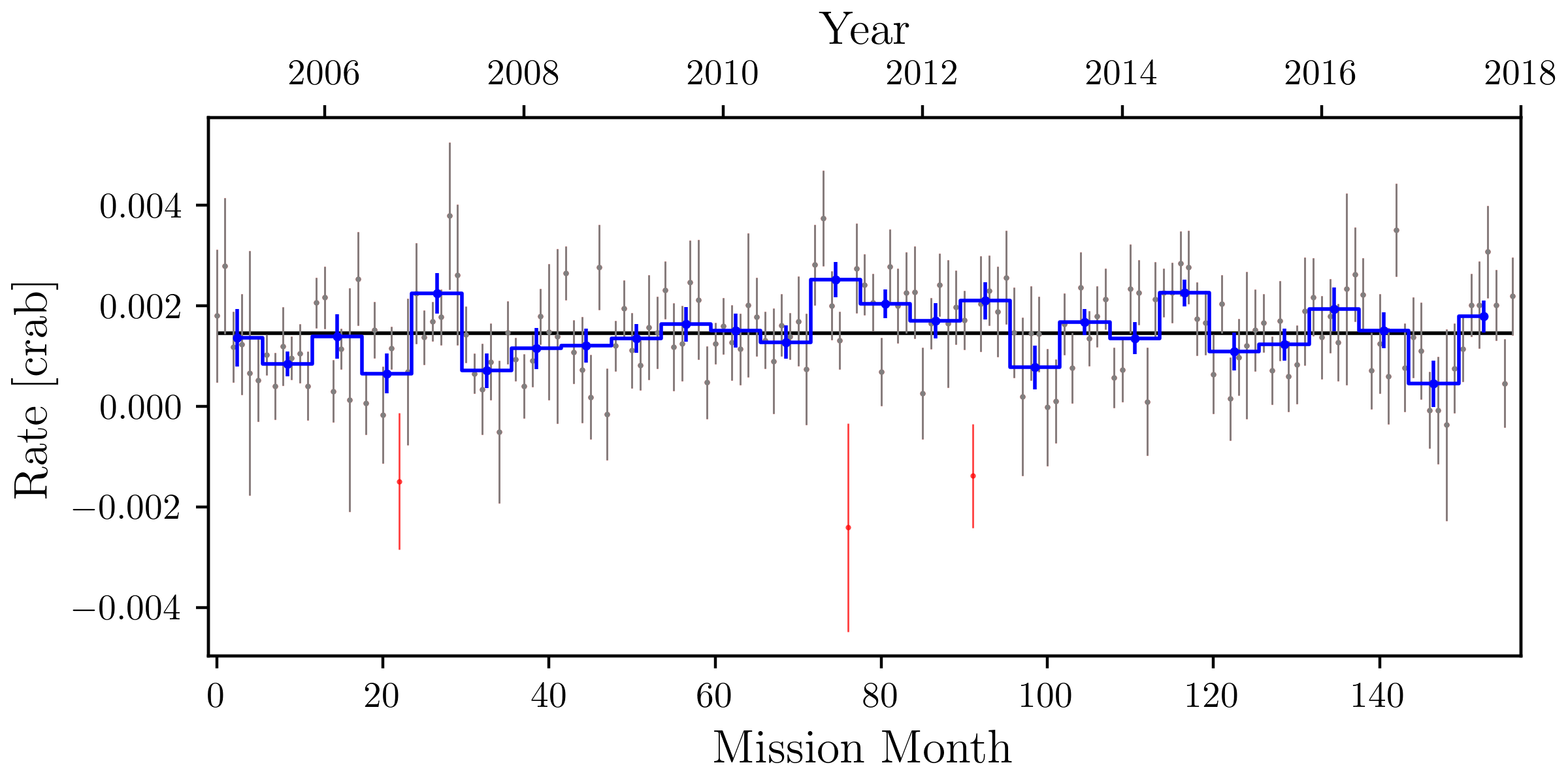}
    \caption
    {BAT-157 Month light curve, the black solid line shows the mean of the 157-month light curve. The grey data are the individual months, while the blue data are the data binned into 6 months. The red data indicates months where the rate does not agree with zero. }
    \label{fig:BAT157}
\end{figure}

In order to accomplish this, we must first attempt to quantify the level of variability from the central engine that is present in \ngc. To avoid the obscuration that affects the soft band, we examine the hard X-ray lightcurve from the BAT-157 month catalogue \citep{BAT+105+month+survay, Lien+2023}. The $14-195$\kev\ band should not be subject to the same level of absorption as the soft band; thus, variation in the central engine should be more easily detected. Figure \ref{fig:BAT157} shows the monthly BAT light curves of \ngc. The red data in this figure indicate months when the source is background-dominated; these data are not used in the analysis but are included here for completeness.   

When compared to the mean count rate, the light curve has a $\chi_{\nu}^2=1.1$ for 153 degrees of freedom. This indicates a relatively constant count rate. Using the \cite{Edelson+2002} method to calculate the fractional variability on this light curve results in a null value due to poor statistics. We bin the data into six mission-month bins to improve the statistics. These data are shown in blue in Figure \ref{fig:BAT157}.  This results in a long-term fractional variability of $27\pm7$\,\% and a $\chi_{\nu}^2=2.3$ when the data is compared to a constant line. This variability provides us with a useful metric to work from. 

To test whether we expect to see any variability in these spectral features with current or future instrumentation, we first assume that the central engine varies by 25 percent on years-long time scales. This is a generous assumption with limited evidence; however, such an assumption must be made. We also assume that all variations in the central engine correspond directly to a change in the luminosity of the emission line component. These assumptions are not exactly based on reality, but instead give us a framework from which we can determine if we could detect variability in the soft band of this source. 

\begin{figure}
    \includegraphics[width=1\linewidth]{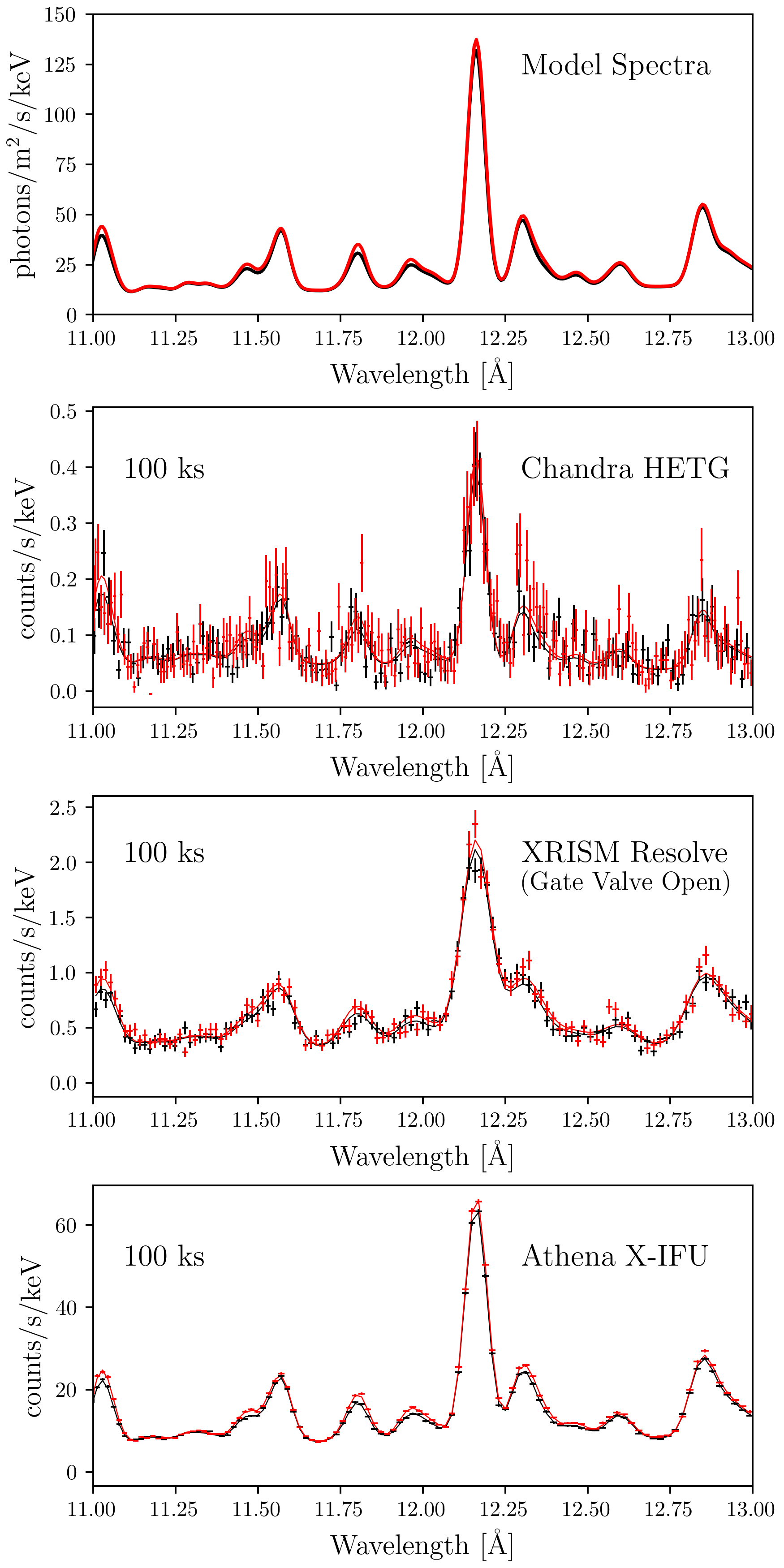}
    \caption
    {Simulations of spectral variability. \textit{Panel A:} the model spectra, the black solid line indicates the base model found in this work. Red indicates the spectrum with the High ionization component  25\% brighter. Panels B, C, and D show simulated data for \Chandra\ HETG, \textit{XRISM} resolve (Gate valve open) and \textit{Athena} X-IFU, respectively, for both the base and 25\% brighter case. All data is simulated for a 100\,ks exposure and all data is binned to three channels per bin.} 
    \label{fig:sims}
\end{figure}

In practice, we adjust the strength of the High ionization component such that its $1-1000$\,Ry luminosity increased by 25 percent. We then simulated 100\,ks of data for both the base model and the 25 percent model using {\sc spex simulate} command. We did this for \Chandra\ HETG, \textit{XRISM} Resolve (assuming gate value open), and \textit{Athena} X-IFU. For all data, we bin by three channels per bin.  The results of these simulations are shown in Figure \ref{fig:sims}. 

The top panel of Figure \ref{fig:sims} shows the expected model for a 25 percent increase in flux for the High ionization component. The difference in the models is minor; the largest change in flux at 11.77\,\AA\ is $\sim15$\,\%. The simulated HETG data shows that if the High ionization component became 25 percent brighter, we would not be able to detect the change with the current data.  We also tested increasing the luminosity by 40 percent in the High ionization component, and we were still unable to differentiate between the base model and the brighter model with HETG. This indicates we are unlikely to detect any intrinsic variability in the emission line components between any epoch of HETG data. Thus, the non-detection of variability is expected and consistent with current data and analysis \citep[e.g.][]{Kallman+2014, Bauer+2015, Grafton+2021}. 

Next, we explore if observation with \textit{XRISM} Resolve \citep{Tashiro+2020, Eckart+2024} of \ngc\ would be able to detect a 25 percent increase in the flux of the High ionization component. We chose to use the gate valve open (GVO) responses as these wavelengths remain inaccessible with the gate valve closed. While the data are of higher quality than the HETG simulations, we would still be unable to detect variation in the emission strength. 
However, a 100\,ks observation of \ngc\ with GVO would produce a spectrum of at least equal quality to the $\sim440\,{\rm ks}$ of HETG data presented here. These data would be helpful in further constraining and confirming the presence of the \fe\ doublet. 

The final simulation is with \textit{Athena} X-IFU \citep{Nandra+2013, Barret+2023}. This is where we may be able to detect flux viability in the High ionization component.  The large collecting area of this instrument would allow for the detection of small changes in the flux of individual lines.  

Once again, the variability observed at higher energy is likely due to changing obscuration along our line of sight \citep[e.g.][]{Marinucci+2016, Zaino+2020}.  It is unclear whether this variability would translate into changes in the strength of emission that is sensitive to radiation through a much larger solid angle.  This section demonstrates that only a deep observation with \textit{Athena} would permit a definitive test.

\subsubsection{High Energy Emission}
Initial indications were that the High ionization component would produce negligible emission compared to the other emission components in the hard energy band. However, this may not be the case. During the modelling of this source, we did not apply any absorption to our ionization model. This is consistent with the previous techniques used to model this source (e.g. \citealt{Kallman+2014, Grafton+2021}) and is common practice for modelling the emission line spectrum of Seyfert 2 galaxies. The fundamental assumption is that the continuum has been completely absorbed and is being re-emitted in an emission line spectrum, \cite{Kinkhabwala+2002} used \ngc\ to demonstrate this. 

While we do not fit the data using any absorption models, we can assume that if \fe\ emission originates from near the inner region, it must be absorbed. To explore this, we applied a fiducial absorber with a column density of $\sim10^{23}\,\cms$ and a covering fraction of $99\%$ to mimic a possible clumpy torus obscuring the line of sight to the High ionization {\sc pion} component. For the emission to be visible with this level of obscuration, it must be intrinsically brighter. We estimate that if this component is intrinsically brighter than the emission from it would be visible in the Fe K$\alpha$ emission region, and therefore might be visible with the \textit{XRISM} observations. Emission lines from Fe have been detected in this region with the HETG and PN instruments \citep{Kallman+2014, Grafton+2021}. However, the detection capabilities of these instruments are limited. HETG lacks the collecting area, and PN lacks the energy resolution to disentangle to complex emission regions that are present in the Fe K$\alpha$ \citep[e.g. NGC\,4151][]{XRISM+2024-NGC4151}. To fully constrain the presence of a plasma this close to the SMBH, broadband modelling that utilizes high-resolution spectroscopy would need to be implemented. This would allow for the simultaneous detection of the $11.77-11.92$\,\AA\ \fe\ lines with any accompanying emission from the  Fe K$\alpha$ band.

\subsection{Other Possible Origins}
\label{sec:ngcOther}
The spectrum of \ngc\ is rich and complex in all energy bands, not just in the X-ray regime. The radio jets in \ngc\ are pushing dense molecular gas ($n_{{\rm H}_2}\sim10^6\, \cmc$) out of the central region \citep{Garcia-Burillo+2014}. Interactions between the jet and the ISM have resulted in a burst of star formation \citep[see][]{Bergeron+1989, Bruhweiler+1991, Melso+2024}. The jet has been shown to ionize material in \ngc\ \citep{Exposito+2011} and in other sources (e.g. NGC\,4388 \citealt{Rod+2017}). Total jet power in the northeastern radio lobe is $P_{jet} \sim7 \times 10^{42}$\,\ergps \citep{Mutie+2024}, which is almost two orders of magnitude greater than the luminosity of the unabsorbed High ionization component. Knots in the radio jet produce much of the $\gamma$-ray emission we see \citep{Salvatore+2024}. High energy emission from the radio jet could be producing a secondary continuum in \ngc\  that is ionizing the material producing the \fe\ emission. This ionizing continuum would be different than those discussed here, so we could not necessarily use the density constraints established in this work.  

Collisionally ionized material may also be responsible for some fraction of this emission. It is very possible in this complex source that these emission lines originate in a collisionally ionized or blended emission region. \cite{Grafton+2021} finds evidence of collisionally ionized emission in the RGS data of this source. The regions of star formation and/or where the jet has interacted with the ISM could be sites of this collisionally ionized emission. The original work by \cite{Mauche+2003} assumes that the \fe\ is ionized by collisional excitations. Thus, if the \fe\ emission originates from a region of collisionally material, we may be able to use the original diagnostic curves, and the density would then be unconstrained (see Section \ref{sec:line}).

We are able to fit the \fe\ region by replacing the High ionization component with one in collisionally ionized equilibrium, {\sc cie}\footnote{Using SPEXACT v3.08.01} \citep{Kaastra+1996}. The fit statistic is comparable to the purely photoionized model. However, the temperature of the plasma is $k{\rm T} = 1.4_{-0.2}^{+0.1}\,\kev$, which is significantly higher than the temperature of the plasma found by \cite{Grafton+2021}. It should be noted that we are fitting only a very narrow region of the spectrum ($11-13$\,\AA). 

The presence of collisionally ionized material would add \ngc\ to a growing list of Seyfert galaxies that have both collisionally and photoionized emission ( e.g., Mrk 573,  \citealt{Paggi+2012}; Mrk 783,  Gonzalez in prep.; Mrk 1239, \citealt{Buhariwalla+2023, Buhariwalla+2024}; NGC 3393, \citealt{Maksym+2019}; NGC 4151, \citealt{Wang+2011}) or absorption (e.g., Mrk 766, \citealt{Matamoro+2025}; NGC 4051, \citealt{Ogorzalek+2022}).

\subsection{Systematic Uncertainty}
The errors calculated in this work have been strictly statistical. They have all been calculated at a 90\% confidence interval. As with any work, some baseline assumptions must be made. Here, we discuss the effects some of these assumptions may have had on our results. 

We expect our photoionization code to be incomplete. 
When \textit{Hitomi} \citep{Takahashi+2016} observed the Perseus cluster, one of the major conclusions was that our photoionized and collisional ionized codes were incomplete \citep{Hitomi+2018B}. In the intervening years, attempts have been made to quantify any systematic errors associated with the atomic data used by line emission codes. 
\cite{Gu+2022} showed that the line flux uncertainties in a collisional plasma could be of order 10\%. While, \cite{Ballhausen+2023}, argue that the majority of systematic effects result from the inadequate model selection in photoionized plasmas and not inadequacies in the atomic data.  

We first assume that our model selection is correct and consider whether uncertainties in our atomic data could reconcile the putative small size ($\sim100\,r_g$) of the \fe\ emitting region. If the measured line value falls below the low-density limit of a diagnostic curve, then we can only determine an upper limit on the density of the emitting material. Thus, the measurement of $100\,r_g$ becomes a lower limit. 

The measured line ratio was $R_{Fe} = 0.41^{+0.17}_{-0.15}$, yielding a 90\% confidence interval of (0.26, 0.58). The line diagnostic in Figure \ref{fig:hiXi} states the lowest value of $I(11.92$\,\AA$)/I(11.77$\,\AA$)$ is $\sim0.17$. Our spectral data (Figure \ref{fig:doublet}) agree with this line ratio within a 99.5\% confidence interval. If we consider the possible flux uncertainty between the emission at 11.77\,\AA\ and 11.92\,\AA\ to be at 10\,\% \citep{Gu+2022}, then the low-density limit could be as high as $\sim0.20$. Our measured ratio agrees with this low-density limit within a 98\,\% confidence interval. This is still greater than a $2\sigma$ result, but it demonstrates how uncertainties in the atomic data may affect the calculated line diagnostic curves. 

Our model assumptions are likely to have an effect as photoionization models are highly dependent on their SED \citep{Mehdipour+2015}. We saw this in Figure \ref{fig:4xsed}, where we presented four different SEDs that could conceivably be created for \ngc. Each one produced a fit that was comparable or superior to our base model. However, the inferred ionization values spanned a full order of magnitude, and the allowed densities ranged from $10^{11} $ to $ 10^{13}\,\cmc$. Notably, none of these four SEDs match those used in previous works. \cite{Kallman+2014} used a power law with $\Gamma=2$, while \cite{Grafton+2021} used a power law, reflection and soft excess components to form their SED. The power law and reflection components were fit to the PN spectrum simultaneously with their pion components.

The common assumption between previous work and our base SED is that all emission regions see the same ionizing continuum as the others. This was, in part, because none of them had any intervening absorbers between the central engine and the emission-line gas. This seems unlikely in such a complex source as \ngc. Having layered absorption is possible with {\sc spex} and can have significant effects on the resultant spectrum (e.g., \citealt{Xu+2025}), but it is very computationally expensive. Furthermore, for a Seyfert 2 galaxy, it would be impossible to ascertain the exact order and properties of the intervening absorbers from the emission line spectra. It is hard enough to identify the base continuum from the central engine in \ngc.

Despite this, we attempted to explore absorption with SED4, where we apply absorption between the ionizing continuum and the emission region. As stated before, this SED produced the best fit to our data between 11 and 13 \AA. The $\Delta {\rm BIC}$ suggests that this model is preferred over the base model with 99.95\% confidence. Furthermore, the ratio of our diagnostic lines in the low-density limit is $I(11.92$\,\AA$)/I(11.77$\,\AA$)=0.25$. This is consistent with the 90\% confidence interval of the measured Fe doublet. The reason density is constrained above the low-density limit may be due to contributions to the \fe\ line from the Mid ionization component. 

We caution against assuming this is the most accurate SED to describe the ionizing continuum that the emission clouds in \ngc\ see. This model only fits over $11-13$\,\AA\ and the warm absorbers are only somewhat grounded in reality. These factors highlight the effects of simply selecting a different SED and do not even consider having different model parameters free (e.g., $\Omega$ vs. $\nh$) or the number of model components.

In summary, the results of the line diagnostic are very model-dependent. Seemingly small changes in the SED result in systematically different diagnostic curves. Furthermore, the ionization state of the plasma has a profound effect on the critical density of the diagnostic curves. As we transition into a high-resolution regime with instruments like \textit{XRISM} and \textit{Athena}, care must be taken in constructing the most realistic SED possible for the photoionized plasma. As well as exercising caution when confronted with exciting or novel parameters.

\section{Conclusions} 
\label{sec:conclusion}
This work explored the density-sensitive \fe\ diagnostic, which has been used to explore the density of emitting plasma surrounding magnetic cataclysmic variable stars. We showed that the diagnostic behaves differently in a photoionized plasma, and calculated the first diagnostic curves for an emission spectrum under an AGN SED. 

We find that the \fe\ doublet is dependent on the ionizing continuum, the ionization state of the plasma, and the density of the material. Thus, it may be difficult to ascertain the density of the emitting plasma if the SED or ionization is unknown. We cannot expect to use the same diagnostic ratio for every AGN and every emitting plasma. Thus, these results remain model-dependent.  These results also highlight the complexity of photoionized emitting plasmas, and \ngc\ in particular. We demonstrate that care must be exercised when interpreting results from spectral models.

Ultimately, it is unlikely that the emitting region for the \fe\ doublet in \ngc\ originates from material $\sim 100\,r_g$ from the central engine. However, this is the result that our photoionized modelling determined under four different SEDs tested. Our spectral fitting demands this result, and we can not exclude this possibility with our variability analysis. Future deep observation \ngc\ with \textit{XRISM} (GVO) and \textit{Athena} are needed to probe the \fe\ lines more fully.

\begin{acknowledgments}
We thank the referee for their thought-provoking comments. This research has made use of data obtained from the Chandra Data Archive provided by the Chandra X-ray Center (CXC). 
\end{acknowledgments}

%

\vspace{5mm}
\facilities{\chandra, \textit{Swift}-BAT}





\bibliography{bibtext}{}

\begin{thebibliography}{}
\expandafter\ifx\csname natexlab\endcsname\relax\def\natexlab#1{#1}\fi
\providecommand{\url}[1]{\href{#1}{#1}}
\providecommand{\dodoi}[1]{doi:~\href{http://doi.org/#1}{\nolinkurl{#1}}}
\providecommand{\doeprint}[1]{\href{http://ascl.net/#1}{\nolinkurl{http://ascl.net/#1}}}
\providecommand{\doarXiv}[1]{\href{https://arxiv.org/abs/#1}{\nolinkurl{https://arxiv.org/abs/#1}}}

\bibitem[{{Arnaud}(1996)}]{Arnaud+1996}
{Arnaud}, K.~A. 1996, in Astronomical Society of the Pacific Conference Series, Vol. 101, Astronomical Data Analysis Software and Systems V, ed. G.~H. {Jacoby} \& J.~{Barnes}, 17

\bibitem[{{Ballhausen} {et~al.}(2023){Ballhausen}, {Kallman}, {Gu}, \& {Paerels}}]{Ballhausen+2023}
{Ballhausen}, R., {Kallman}, T.~R., {Gu}, L., \& {Paerels}, F. 2023, \apj, 956, 65, \dodoi{10.3847/1538-4357/aced49}

\bibitem[{{Barret} {et~al.}(2023){Barret}, {Albouys}, {Herder}, {Piro}, {Cappi}, {Huovelin}, {Kelley}, {Mas-Hesse}, {Paltani}, {Rauw}, {Rozanska}, {Svoboda}, {Wilms}, {Yamasaki}, {Audard}, {Bandler}, {Barbera}, {Barcons}, {Bozzo}, {Ceballos}, {Charles}, {Costantini}, {Dauser}, {Decourchelle}, {Duband}, {Duval}, {Fiore}, {Gatti}, {Goldwurm}, {Hartog}, {Jackson}, {Jonker}, {Kilbourne}, {Korpela}, {Macculi}, {Mendez}, {Mitsuda}, {Molendi}, {Pajot}, {Pointecouteau}, {Porter}, {Pratt}, {Pr{\^e}le}, {Ravera}, {Sato}, {Schaye}, {Shinozaki}, {Skup}, {Soucek}, {Thibert}, {Vink}, {Webb}, {Chaoul}, {Raulin}, {Simionescu}, {Torrejon}, {Acero}, {Branduardi-Raymont}, {Ettori}, {Finoguenov}, {Grosso}, {Kaastra}, {Mazzotta}, {Miller}, {Miniutti}, {Nicastro}, {Sciortino}, {Yamaguchi}, {Beaumont}, {Cucchetti}, {D'Andrea}, {Eckart}, {Ferrando}, {Kammoun}, {Lotti}, {Mesnager}, {Natalucci}, {Peille}, {de Plaa}, {Ardellier}, {Argan}, {Bellouard}, {Carron}, {Cavazzuti}, {Fiorini}, {Khosropanah}, {Martin}, {Perry}, {Pinsard},
  {Pradines}, {Rigano}, {Roelfsema}, {Schwander}, {Torrioli}, {Ullom}, {Vera}, {Villegas}, {Zuchniak}, {Brachet}, {Cicero}, {Doriese}, {Durkin}, {Fioretti}, {Geoffray}, {Jacques}, {Kirsch}, {Smith}, {Adams}, {Gloaguen}, {Hoogeveen}, {van der Hulst}, {Kiviranta}, {van der Kuur}, {Ledot}, {van Leeuwen}, {van Loon}, {Lyautey}, {Parot}, {Sakai}, {van Weers}, {Abdoelkariem}, {Adam}, {Adami}, {Aicardi}, {Akamatsu}, {Alonso}, {Amato}, {Andr{\'e}}, {Angelinelli}, {Anon-Cancela}, {Anvar}, {Atienza}, {Attard}, {Auricchio}, {Balado}, {Bancel}, {Barusso}, {Bascu{\~n}an}, {Bernard}, {Berrocal}, {Blin}, {Bonino}, {Bonnet}, {Bonny}, {Boorman}, {Boreux}, {Bounab}, {Boutelier}, {Boyce}, {Brienza}, {Bruijn}, {Bulgarelli}, {Calarco}, {Callanan}, {Campello}, {Camus}, {Canourgues}, {Capobianco}, {Cardiel}, {Castellani}, {Cheatom}, {Chervenak}, {Chiarello}, {Clerc}, {Clerc}, {Cobo}, {Coeur-Joly}, {Coleiro}, {Colonges}, {Corcione}, {Coriat}, {Coynel}, {Cuttaia}, {D'Ai}, {D'anca}, {Dadina}, {Daniel}, {Dauner}, {DeNigris},
  {Dercksen}, {DiPirro}, {Doumayrou}, {Dubbeldam}, {Dupieux}, {Dupourqu{\'e}}, {Durand}, {Eckert}, {Eiriz}, {Ercolani}, {Etcheverry}, {Finkbeiner}, {Fiocchi}, {Fossecave}, {Franssen}, {Frericks}, {Gabici}, {Gant}, {Gao}, {Gastaldello}, \& {Genolet}}]{Barret+2023}
{Barret}, D., {Albouys}, V., {Herder}, J.-W.~d., {et~al.} 2023, Experimental Astronomy, 55, 373, \dodoi{10.1007/s10686-022-09880-7}

\bibitem[{{Bauer} {et~al.}(2015){Bauer}, {Ar{\'e}valo}, {Walton}, {Koss}, {Puccetti}, {Gandhi}, {Stern}, {Alexander}, {Balokovi{\'c}}, {Boggs}, {Brandt}, {Brightman}, {Christensen}, {Comastri}, {Craig}, {Del Moro}, {Hailey}, {Harrison}, {Hickox}, {Luo}, {Markwardt}, {Marinucci}, {Matt}, {Rigby}, {Rivers}, {Saez}, {Treister}, {Urry}, \& {Zhang}}]{Bauer+2015}
{Bauer}, F.~E., {Ar{\'e}valo}, P., {Walton}, D.~J., {et~al.} 2015, \apj, 812, 116, \dodoi{10.1088/0004-637X/812/2/116}

\bibitem[{{Bergeron} {et~al.}(1989){Bergeron}, {Petitjean}, \& {Durret}}]{Bergeron+1989}
{Bergeron}, J., {Petitjean}, P., \& {Durret}, F. 1989, \aap, 213, 61

\bibitem[{{Bland-Hawthorn} {et~al.}(1997){Bland-Hawthorn}, {Gallimore}, {Tacconi}, {Brinks}, {Baum}, {Antonucci}, \& {Cecil}}]{Bland-Hawthorn+1997}
{Bland-Hawthorn}, J., {Gallimore}, J.~F., {Tacconi}, L.~J., {et~al.} 1997, \apss, 248, 9, \dodoi{10.1023/A:1000567831370}

\bibitem[{{Blustin} {et~al.}(2005){Blustin}, {Page}, {Fuerst}, {Branduardi-Raymont}, \& {Ashton}}]{Blustin+2005}
{Blustin}, A.~J., {Page}, M.~J., {Fuerst}, S.~V., {Branduardi-Raymont}, G., \& {Ashton}, C.~E. 2005, \aap, 431, 111, \dodoi{10.1051/0004-6361:20041775}

\bibitem[{{Bruhweiler} {et~al.}(1991){Bruhweiler}, {Truong}, \& {Altner}}]{Bruhweiler+1991}
{Bruhweiler}, F.~C., {Truong}, K.~Q., \& {Altner}, B. 1991, \apj, 379, 596, \dodoi{10.1086/170532}

\bibitem[{{Buhariwalla} {et~al.}(2024){Buhariwalla}, {Gallo}, {Mao}, {Jiang}, {Pothier-Bogoslowski}, {J{\"a}rvel{\"a}}, {Komossa}, \& {Grupe}}]{Buhariwalla+2024}
{Buhariwalla}, M.~Z., {Gallo}, L.~C., {Mao}, J., {et~al.} 2024, \apj, 971, 22, \dodoi{10.3847/1538-4357/ad4ee0}

\bibitem[{{Buhariwalla} {et~al.}(2023){Buhariwalla}, {Gallo}, {Mao}, {Komossa}, {Jiang}, {Gonzalez}, \& {Grupe}}]{Buhariwalla+2023}
---. 2023, \mnras, 521, 2378, \dodoi{10.1093/mnras/stad265}

\bibitem[{{Burke} {et~al.}(2023){Burke}, {Laurino}, {wmclaugh}, {G{\"u}nther}, {Marie-Terrell}, {dtnguyen2}, {Siemiginowska}, {Cheer}, {Budynkiewicz}, {Aldcroft}, {Deil}, {Sip{\H{o}}cz}, {Buchner}, {nplee}, {Donath}, {Laginja}, {Leinweber}, \& {Todd}}]{Burke+2023}
{Burke}, D., {Laurino}, O., {wmclaugh}, {et~al.} 2023, {sherpa/sherpa: Sherpa 4.16.0}, 4.16.0,  Zenodo, \dodoi{10.5281/zenodo.825839}

\bibitem[{{Canizares} {et~al.}(2000){Canizares}, {Huenemoerder}, {Davis}, {Dewey}, {Flanagan}, {Houck}, {Markert}, {Marshall}, {Schattenburg}, {Schulz}, {Wise}, {Drake}, \& {Brickhouse}}]{Canizares+2000}
{Canizares}, C.~R., {Huenemoerder}, D.~P., {Davis}, D.~S., {et~al.} 2000, \apjl, 539, L41, \dodoi{10.1086/312823}

\bibitem[{{Chen} {et~al.}(2004){Chen}, {Beiersdorfer}, {Heeter}, {Liedahl}, {Naranjo-Rivera}, {Tr{\"a}bert}, {Gu}, \& {Lepson}}]{Chen+2004}
{Chen}, H., {Beiersdorfer}, P., {Heeter}, L.~A., {et~al.} 2004, \apj, 611, 598, \dodoi{10.1086/421987}

\bibitem[{{Di Gesu} {et~al.}(2017){Di Gesu}, {Costantini}, {Piconcelli}, {Kaastra}, {Mehdipour}, \& {Paltani}}]{DiGesu+2017}
{Di Gesu}, L., {Costantini}, E., {Piconcelli}, E., {et~al.} 2017, \aap, 608, A115, \dodoi{10.1051/0004-6361/201731853}

\bibitem[{{Done} {et~al.}(2012){Done}, {Davis}, {Jin}, {Blaes}, \& {Ward}}]{Done+2012}
{Done}, C., {Davis}, S.~W., {Jin}, C., {Blaes}, O., \& {Ward}, M. 2012, \mnras, 420, 1848, \dodoi{10.1111/j.1365-2966.2011.19779.x}

\bibitem[{Eckart {et~al.}(2024)Eckart, Brown, Chiao, Cumbee, Fujimoto, Hell, Hoshino, Ishisaki, Kelley, Kenyon, Kilbourne, Kitamoto, Leutenegger, Lockard, Loewenstein, Magee, Mizumoto, Porter, Sato, Sawada, Shah, Shipman, Sneiderman, Takei, Tsujimoto, de~Vries, Watanabe, Witthoeft, Wolfs, Yamada, \& Yaqoob}]{Eckart+2024}
Eckart, M.~E., Brown, G.~V., Chiao, M.~P., {et~al.} 2024, in Space Telescopes and Instrumentation 2024: Ultraviolet to Gamma Ray, ed. J.-W.~A. den Herder, S.~Nikzad, \& K.~Nakazawa, Vol. 13093, International Society for Optics and Photonics (SPIE), 130931P, \dodoi{10.1117/12.3019276}

\bibitem[{Edelson {et~al.}(2002)Edelson, Turner, Pounds, Vaughan, Markowitz, Marshall, Dobbie, \& Warwick}]{Edelson+2002}
Edelson, R., Turner, T.~J., Pounds, K., {et~al.} 2002, \apj, 568, 610, \dodoi{10.1086/323779}

\bibitem[{{Exposito} {et~al.}(2011){Exposito}, {Gratadour}, {Cl{\'e}net}, \& {Rouan}}]{Exposito+2011}
{Exposito}, J., {Gratadour}, D., {Cl{\'e}net}, Y., \& {Rouan}, D. 2011, \aap, 533, A63, \dodoi{10.1051/0004-6361/201116943}

\bibitem[{{Fabian}(1994)}]{Fabian+1994}
{Fabian}, A.~C. 1994, \araa, 32, 277, \dodoi{10.1146/annurev.aa.32.090194.001425}

\bibitem[{{Fruscione} {et~al.}(2006){Fruscione}, {McDowell}, {Allen}, {Brickhouse}, {Burke}, {Davis}, {Durham}, {Elvis}, {Galle}, {Harris}, {Huenemoerder}, {Houck}, {Ishibashi}, {Karovska}, {Nicastro}, {Noble}, {Nowak}, {Primini}, {Siemiginowska}, {Smith}, \& {Wise}}]{Fruscione+2006}
{Fruscione}, A., {McDowell}, J.~C., {Allen}, G.~E., {et~al.} 2006, in Society of Photo-Optical Instrumentation Engineers (SPIE) Conference Series, Vol. 6270, Observatory Operations: Strategies, Processes, and Systems, ed. D.~R. {Silva} \& R.~E. {Doxsey}, 62701V, \dodoi{10.1117/12.671760}

\bibitem[{{Gallimore} \& {Impellizzeri}(2023)}]{Gallimore+2023}
{Gallimore}, J.~F., \& {Impellizzeri}, C.~M.~V. 2023, \apj, 951, 109, \dodoi{10.3847/1538-4357/acd846}

\bibitem[{{Gallo} {et~al.}(2023){Gallo}, {Miller}, \& {Costantini}}]{Gallo+2023}
{Gallo}, L.~C., {Miller}, J.~M., \& {Costantini}, E. 2023, arXiv e-prints, arXiv:2302.10930, \dodoi{10.48550/arXiv.2302.10930}

\bibitem[{{Garc{\'\i}a-Burillo} {et~al.}(2014){Garc{\'\i}a-Burillo}, {Combes}, {Usero}, {Aalto}, {Krips}, {Viti}, {Alonso-Herrero}, {Hunt}, {Schinnerer}, {Baker}, {Boone}, {Casasola}, {Colina}, {Costagliola}, {Eckart}, {Fuente}, {Henkel}, {Labiano}, {Mart{\'\i}n}, {M{\'a}rquez}, {Muller}, {Planesas}, {Ramos Almeida}, {Spaans}, {Tacconi}, \& {van der Werf}}]{Garcia-Burillo+2014}
{Garc{\'\i}a-Burillo}, S., {Combes}, F., {Usero}, A., {et~al.} 2014, \aap, 567, A125, \dodoi{10.1051/0004-6361/201423843}

\bibitem[{{Grafton-Waters} {et~al.}(2021){Grafton-Waters}, {Branduardi-Raymont}, {Mehdipour}, {Page}, {Bianchi}, {Behar}, \& {Symeonidis}}]{Grafton+2021}
{Grafton-Waters}, S., {Branduardi-Raymont}, G., {Mehdipour}, M., {et~al.} 2021, \aap, 649, A162, \dodoi{10.1051/0004-6361/202039022}

\bibitem[{{Grafton-Waters} {et~al.}(2023){Grafton-Waters}, {Mao, J.}, {Mehdipour, M.}, {Branduardi-Raymont, G.}, {Page, M.}, {Kaastra, J.}, {Wang, Y.}, {Pinto, C.}, {Kriss, G. A.}, {Walton, D. J.}, {Petrucci, P.-O.}, {Ponti, G.}, {De Marco, B.}, {Bianchi, S.}, {Behar, E.}, \& {Ebrero, J.}}]{Grafton+2023}
{Grafton-Waters}, S., {Mao, J.}, {Mehdipour, M.}, {et~al.} 2023, \aap, 673, A26, \dodoi{10.1051/0004-6361/202243681}

\bibitem[{{Greenhill} \& {Gwinn}(1997)}]{Greenhill+1997}
{Greenhill}, L.~J., \& {Gwinn}, C.~R. 1997, \apss, 248, 261, \dodoi{10.1023/A:1000554317683}

\bibitem[{{Grupe} {et~al.}(2001){Grupe}, {Thomas}, \& {Beuermann}}]{Grupe+2001}
{Grupe}, D., {Thomas}, H.~C., \& {Beuermann}, K. 2001, \aap, 367, 470, \dodoi{10.1051/0004-6361:20000429}

\bibitem[{{Gu} {et~al.}(2022){Gu}, {Shah}, {Mao}, {Raassen}, {de Plaa}, {Pinto}, {Akamatsu}, {Werner}, {Simionescu}, {Mernier}, {Sawada}, {Mohanty}, {Amaro}, {Gu}, {Porter}, {L{\'o}pez-Urrutia}, \& {Kaastra}}]{Gu+2022}
{Gu}, L., {Shah}, C., {Mao}, J., {et~al.} 2022, \aap, 664, A62, \dodoi{10.1051/0004-6361/202039943}

\bibitem[{{Hitomi Collaboration} {et~al.}(2018){Hitomi Collaboration}, {Aharonian}, {Akamatsu}, {Akimoto}, {Allen}, {Angelini}, {Audard}, {Awaki}, {Axelsson}, {Bamba}, {Bautz}, {Blandford}, {Brenneman}, {Brown}, {Bulbul}, {Cackett}, {Chernyakova}, {Chiao}, {Coppi}, {Costantini}, {de Plaa}, {de Vries}, {den Herder}, {Done}, {Dotani}, {Ebisawa}, {Eckart}, {Enoto}, {Ezoe}, {Fabian}, {Ferrigno}, {Foster}, {Fujimoto}, {Fukazawa}, {Furuzawa}, {Galeazzi}, {Gallo}, {Gandhi}, {Giustini}, {Goldwurm}, {Gu}, {Guainazzi}, {Haba}, {Hagino}, {Hamaguchi}, {Harrus}, {Hatsukade}, {Hayashi}, {Hayashi}, {Hayashida}, {Hell}, {Hiraga}, {Hornschemeier}, {Hoshino}, {Hughes}, {Ichinohe}, {Iizuka}, {Inoue}, {Inoue}, {Ishida}, {Ishikawa}, {Ishisaki}, {Iwai}, {Kaastra}, {Kallman}, {Kamae}, {Kataoka}, {Katsuda}, {Kawai}, {Kelley}, {Kilbourne}, {Kitaguchi}, {Kitamoto}, {Kitayama}, {Kohmura}, {Kokubun}, {Koyama}, {Koyama}, {Kretschmar}, {Krimm}, {Kubota}, {Kunieda}, {Laurent}, {Lee}, {Leutenegger}, {Limousin}, {Loewenstein}, {Long},
  {Lumb}, {Madejski}, {Maeda}, {Maier}, {Makishima}, {Markevitch}, {Matsumoto}, {Matsushita}, {McCammon}, {McNamara}, {Mehdipour}, {Miller}, {Miller}, {Mineshige}, {Mitsuda}, {Mitsuishi}, {Miyazawa}, {Mizuno}, {Mori}, {Mori}, {Mukai}, {Murakami}, {Mushotzky}, {Nakagawa}, {Nakajima}, {Nakamori}, {Nakashima}, {Nakazawa}, {Nobukawa}, {Nobukawa}, {Noda}, {Odaka}, {Ohashi}, {Ohno}, {Okajima}, {Ota}, {Ozaki}, {Paerels}, {Paltani}, {Petre}, {Pinto}, {Porter}, {Pottschmidt}, {Reynolds}, {Safi-Harb}, {Saito}, {Sakai}, {Sasaki}, {Sato}, {Sato}, {Sato}, {Sawada}, {Schartel}, {Serlemtsos}, {Seta}, {Shidatsu}, {Simionescu}, {Smith}, {Soong}, {Stawarz}, {Sugawara}, {Sugita}, {Szymkowiak}, {Tajima}, {Takahashi}, {Takahashi}, {Takeda}, {Takei}, {Tamagawa}, {Tamura}, {Tanaka}, {Tanaka}, {Tanaka}, {Tashiro}, {Tawara}, {Terada}, {Terashima}, {Tombesi}, {Tomida}, {Tsuboi}, {Tsujimoto}, {Tsunemi}, {Tsuru}, {Uchida}, {Uchiyama}, {Uchiyama}, {Ueda}, {Ueda}, {Uno}, {Urry}, {Ursino}, {Watanabe}, {Werner}, {Wilkins}, {Williams},
  {Yamada}, {Yamaguchi}, {Yamaoka}, {Yamasaki}, {Yamauchi}, {Yamauchi}, {Yaqoob}, {Yatsu}, {Yonetoku}, {Zhuravleva}, {Zoghbi}, \& {Raassen}}]{Hitomi+2018B}
{Hitomi Collaboration}, {Aharonian}, F., {Akamatsu}, H., {et~al.} 2018, \pasj, 70, 12, \dodoi{10.1093/pasj/psx156}

\bibitem[{{Huchra} {et~al.}(1999){Huchra}, {Vogeley}, \& {Geller}}]{Huchra+1999}
{Huchra}, J.~P., {Vogeley}, M.~S., \& {Geller}, M.~J. 1999, \apjs, 121, 287, \dodoi{10.1086/313194}

\bibitem[{{Huenemoerder} {et~al.}(2011){Huenemoerder}, {Mitschang}, {Dewey}, {Nowak}, {Schulz}, {Nichols}, {Davis}, {Houck}, {Marshall}, {Noble}, {Morgan}, \& {Canizares}}]{Huenemoerder+2011}
{Huenemoerder}, D.~P., {Mitschang}, A., {Dewey}, D., {et~al.} 2011, \aj, 141, 129, \dodoi{10.1088/0004-6256/141/4/129}

\bibitem[{{Kaastra}(2017)}]{Kaastra+2017}
{Kaastra}, J.~S. 2017, \aap, 605, A51, \dodoi{10.1051/0004-6361/201629319}

\bibitem[{{Kaastra} \& {Bleeker}(2016)}]{Kaastra+2016}
{Kaastra}, J.~S., \& {Bleeker}, J. A.~M. 2016, A\&A, 587, A151, \dodoi{10.1051/0004-6361/201527395}

\bibitem[{{Kaastra} {et~al.}(1996){Kaastra}, {Mewe}, \& {Nieuwenhuijzen}}]{Kaastra+1996}
{Kaastra}, J.~S., {Mewe}, R., \& {Nieuwenhuijzen}, H. 1996, in UV and X-ray Spectroscopy of Astrophysical and Laboratory Plasmas, ed. K.~{Yamashita} \& T.~{Watanabe}, 411--414

\bibitem[{{Kaastra} {et~al.}(2024){Kaastra}, {Raassen}, {de Plaa}, \& {Gu}}]{Kaastra+2024}
{Kaastra}, J.~S., {Raassen}, A.~J.~J., {de Plaa}, J., \& {Gu}, L. 2024, {SPEX X-ray spectral fitting package}, 3.08.00,  Zenodo, \dodoi{10.5281/zenodo.10822753}

\bibitem[{{Kallman} {et~al.}(2014){Kallman}, {Evans}, {Marshall}, {Canizares}, {Longinotti}, {Nowak}, \& {Schulz}}]{Kallman+2014}
{Kallman}, T., {Evans}, D.~A., {Marshall}, H., {et~al.} 2014, \apj, 780, 121, \dodoi{10.1088/0004-637X/780/2/121}

\bibitem[{{Kallman} \& {McCray}(1982)}]{Kallman+1982}
{Kallman}, T.~R., \& {McCray}, R. 1982, \apjs, 50, 263, \dodoi{10.1086/190828}

\bibitem[{Kass \& Raftery(1995)}]{Kass+1995}
Kass, R.~E., \& Raftery, A.~E. 1995, Journal of the American Statistical Association, 90, 773, \dodoi{10.1080/01621459.1995.10476572}

\bibitem[{{King} {et~al.}(2012){King}, {Miller}, \& {Raymond}}]{King+2012}
{King}, A.~L., {Miller}, J.~M., \& {Raymond}, J. 2012, \apj, 746, 2, \dodoi{10.1088/0004-637X/746/1/2}

\bibitem[{{Kinkhabwala} {et~al.}(2002){Kinkhabwala}, {Sako}, {Behar}, {Kahn}, {Paerels}, {Brinkman}, {Kaastra}, {Gu}, \& {Liedahl}}]{Kinkhabwala+2002}
{Kinkhabwala}, A., {Sako}, M., {Behar}, E., {et~al.} 2002, \apj, 575, 732, \dodoi{10.1086/341482}

\bibitem[{Kramida {et~al.}(2024)Kramida, {Yu.~Ralchenko}, Reader, \& {and NIST ASD Team}}]{NIST+2024}
Kramida, A., {Yu.~Ralchenko}, Reader, J., \& {and NIST ASD Team}. 2024, {NIST Atomic Spectra Database (ver. 5.12), [Online]. Available: {\tt{https://physics.nist.gov/asd}} [2025, April 15]. National Institute of Standards and Technology, Gaithersburg, MD.}

\bibitem[{{Kubota} \& {Done}(2018)}]{Kubota+2018}
{Kubota}, A., \& {Done}, C. 2018, \mnras, 480, 1247, \dodoi{10.1093/mnras/sty1890}

\bibitem[{{Laha} {et~al.}(2014){Laha}, {Guainazzi}, {Dewangan}, {Chakravorty}, \& {Kembhavi}}]{Laha+2014}
{Laha}, S., {Guainazzi}, M., {Dewangan}, G.~C., {Chakravorty}, S., \& {Kembhavi}, A.~K. 2014, \mnras, 441, 2613, \dodoi{10.1093/mnras/stu669}

\bibitem[{Lange(2023)}]{nautilus}
Lange, J.~U. 2023, Monthly Notices of the Royal Astronomical Society, 525, 3181, \dodoi{10.1093/mnras/stad2441}

\bibitem[{{Leighly}(1999{\natexlab{a}})}]{Leighly+1999A}
{Leighly}, K.~M. 1999{\natexlab{a}}, \apjs, 125, 297, \dodoi{10.1086/313277}

\bibitem[{{Leighly}(1999{\natexlab{b}})}]{Leighly+1999B}
---. 1999{\natexlab{b}}, ApJs, 125, 317, \dodoi{10.1086/313287}

\bibitem[{{Liang} \& {Zhao}(2008)}]{Liang+2008}
{Liang}, G.~Y., \& {Zhao}, G. 2008, \aj, 135, 2291, \dodoi{10.1088/0004-6256/135/6/2291}

\bibitem[{{Lien} {et~al.}(2023){Lien}, {Krimm}, {Markwardt}, {Collins}, {Barthelmy}, {Oh}, {Koss}, {Parsotan}, \& {Cenko}}]{Lien+2023}
{Lien}, A., {Krimm}, H., {Markwardt}, C., {et~al.} 2023, in American Astronomical Society Meeting Abstracts, Vol. 241, American Astronomical Society Meeting Abstracts, 254.07

\bibitem[{{Lodato} \& {Bertin}(2003)}]{Lodato+2003}
{Lodato}, G., \& {Bertin}, G. 2003, \aap, 398, 517, \dodoi{10.1051/0004-6361:20021672}

\bibitem[{{Lusso} {et~al.}(2010){Lusso}, {Comastri}, {Vignali}, {Zamorani}, {Brusa}, {Gilli}, {Iwasawa}, {Salvato}, {Civano}, {Elvis}, {Merloni}, {Bongiorno}, {Trump}, {Koekemoer}, {Schinnerer}, {Le Floc'h}, {Cappelluti}, {Jahnke}, {Sargent}, {Silverman}, {Mainieri}, {Fiore}, {Bolzonella}, {Le F{\`e}vre}, {Garilli}, {Iovino}, {Kneib}, {Lamareille}, {Lilly}, {Mignoli}, {Scodeggio}, \& {Vergani}}]{Lusso+2010}
{Lusso}, E., {Comastri}, A., {Vignali}, C., {et~al.} 2010, \aap, 512, A34, \dodoi{10.1051/0004-6361/200913298}

\bibitem[{Maksym {et~al.}(2019)Maksym, Fabbiano, Elvis, Karovska, Paggi, Raymond, Wang, Storchi-Bergmann, \& Risaliti}]{Maksym+2019}
Maksym, W.~P., Fabbiano, G., Elvis, M., {et~al.} 2019, The Astrophysical Journal, 872, 94, \dodoi{10.3847/1538-4357/aaf4f5}

\bibitem[{{Mao} {et~al.}(2017){Mao}, {Kaastra}, {Mehdipour}, {Raassen}, {Gu}, \& {Miller}}]{Mao+2017}
{Mao}, J., {Kaastra}, J.~S., {Mehdipour}, M., {et~al.} 2017, \aap, 607, A100, \dodoi{10.1051/0004-6361/201731378}

\bibitem[{{Marinucci} {et~al.}(2016){Marinucci}, {Bianchi}, {Matt}, {Alexander}, {Balokovi{\'c}}, {Bauer}, {Brandt}, {Gandhi}, {Guainazzi}, {Harrison}, {Iwasawa}, {Koss}, {Madsen}, {Nicastro}, {Puccetti}, {Ricci}, {Stern}, \& {Walton}}]{Marinucci+2016}
{Marinucci}, A., {Bianchi}, S., {Matt}, G., {et~al.} 2016, \mnras, 456, L94, \dodoi{10.1093/mnrasl/slv178}

\bibitem[{{Matamoro Zatarain} {et~al.}(2025){Matamoro Zatarain}, {Costantini}, {Jur{\'a}{\v{n}}ov{\'a}}, \& {Rogantini}}]{Matamoro+2025}
{Matamoro Zatarain}, T., {Costantini}, E., {Jur{\'a}{\v{n}}ov{\'a}}, A., \& {Rogantini}, D. 2025, arXiv e-prints, arXiv:2501.11562, \dodoi{10.48550/arXiv.2501.11562}

\bibitem[{{Mauche} {et~al.}(2003){Mauche}, {Liedahl}, \& {Fournier}}]{Mauche+2003}
{Mauche}, C.~W., {Liedahl}, D.~A., \& {Fournier}, K.~B. 2003, \apjl, 588, L101, \dodoi{10.1086/375684}

\bibitem[{{Mauche} {et~al.}(2004){Mauche}, {Liedahl}, \& {Fournier}}]{Mauche+2004}
{Mauche}, C.~W., {Liedahl}, D.~A., \& {Fournier}, K.~B. 2004, in Astronomical Society of the Pacific Conference Series, Vol. 315, IAU Colloq. 190: Magnetic Cataclysmic Variables, ed. S.~{Vrielmann} \& M.~{Cropper}, 124, \dodoi{10.48550/arXiv.astro-ph/0301633}

\bibitem[{{Mehdipour} {et~al.}(2016){Mehdipour}, {Kaastra}, \& {Kallman}}]{Mehdipour+2016}
{Mehdipour}, M., {Kaastra}, J.~S., \& {Kallman}, T. 2016, \aap, 596, A65, \dodoi{10.1051/0004-6361/201628721}

\bibitem[{{Mehdipour} {et~al.}(2015){Mehdipour}, {Kaastra}, {Kriss}, {Cappi}, {Petrucci}, {Steenbrugge}, {Arav}, {Behar}, {Bianchi}, {Boissay}, {Branduardi-Raymont}, {Costantini}, {Ebrero}, {Di Gesu}, {Harrison}, {Kaspi}, {De Marco}, {Matt}, {Paltani}, {Peterson}, {Ponti}, {Pozo Nu{\~n}ez}, {De Rosa}, {Ursini}, {de Vries}, {Walton}, \& {Whewell}}]{Mehdipour+2015}
{Mehdipour}, M., {Kaastra}, J.~S., {Kriss}, G.~A., {et~al.} 2015, \aap, 575, A22, \dodoi{10.1051/0004-6361/201425373}

\bibitem[{{Melso} {et~al.}(2024){Melso}, {Schiminovich}, {Sitaram}, {Cevallos-Aleman}, {Cruvinel Santiago}, {Smiley}, \& {Ong}}]{Melso+2024}
{Melso}, N., {Schiminovich}, D., {Sitaram}, M., {et~al.} 2024, \apj, 974, 161, \dodoi{10.3847/1538-4357/ad6cd1}

\bibitem[{{Miller} {et~al.}(2014){Miller}, {Raymond}, {Kallman}, {Maitra}, {Fabian}, {Proga}, {Reynolds}, {Reynolds}, {Degenaar}, {King}, {Cackett}, {Kennea}, \& {Beardmore}}]{Miller+2014}
{Miller}, J.~M., {Raymond}, J., {Kallman}, T.~R., {et~al.} 2014, \apj, 788, 53, \dodoi{10.1088/0004-637X/788/1/53}

\bibitem[{{Miller} {et~al.}(2015){Miller}, {Kaastra}, {Miller}, {Reynolds}, {Brown}, {Cenko}, {Drake}, {Gezari}, {Guillochon}, {Gultekin}, {Irwin}, {Levan}, {Maitra}, {Maksym}, {Mushotzky}, {O'Brien}, {Paerels}, {de Plaa}, {Ramirez-Ruiz}, {Strohmayer}, \& {Tanvir}}]{Miller+2015}
{Miller}, J.~M., {Kaastra}, J.~S., {Miller}, M.~C., {et~al.} 2015, \nat, 526, 542, \dodoi{10.1038/nature15708}

\bibitem[{{Mutie} {et~al.}(2024){Mutie}, {Williams-Baldwin}, {Beswick}, {Bempong-Manful}, {Baki}, {Muxlow}, {Gallimore}, {Aalto}, {Dullo}, \& {Baldi}}]{Mutie+2024}
{Mutie}, I.~M., {Williams-Baldwin}, D., {Beswick}, R.~J., {et~al.} 2024, \mnras, 527, 11756, \dodoi{10.1093/mnras/stad3864}

\bibitem[{{Nandra} {et~al.}(2013){Nandra}, {Barret}, {Barcons}, {Fabian}, {den Herder}, {Piro}, {Watson}, {Adami}, {Aird}, {Afonso}, {Alexander}, {Argiroffi}, {Amati}, {Arnaud}, {Atteia}, {Audard}, {Badenes}, {Ballet}, {Ballo}, {Bamba}, {Bhardwaj}, {Stefano Battistelli}, {Becker}, {De Becker}, {Behar}, {Bianchi}, {Biffi}, {B{\^\i}rzan}, {Bocchino}, {Bogdanov}, {Boirin}, {Boller}, {Borgani}, {Borm}, {Bouch{\'e}}, {Bourdin}, {Bower}, {Braito}, {Branchini}, {Branduardi-Raymont}, {Bregman}, {Brenneman}, {Brightman}, {Br{\"u}ggen}, {Buchner}, {Bulbul}, {Brusa}, {Bursa}, {Caccianiga}, {Cackett}, {Campana}, {Cappelluti}, {Cappi}, {Carrera}, {Ceballos}, {Christensen}, {Chu}, {Churazov}, {Clerc}, {Corbel}, {Corral}, {Comastri}, {Costantini}, {Croston}, {Dadina}, {D'Ai}, {Decourchelle}, {Della Ceca}, {Dennerl}, {Dolag}, {Done}, {Dovciak}, {Drake}, {Eckert}, {Edge}, {Ettori}, {Ezoe}, {Feigelson}, {Fender}, {Feruglio}, {Finoguenov}, {Fiore}, {Galeazzi}, {Gallagher}, {Gandhi}, {Gaspari}, {Gastaldello}, {Georgakakis},
  {Georgantopoulos}, {Gilfanov}, {Gitti}, {Gladstone}, {Goosmann}, {Gosset}, {Grosso}, {Guedel}, {Guerrero}, {Haberl}, {Hardcastle}, {Heinz}, {Alonso Herrero}, {Herv{\'e}}, {Holmstrom}, {Iwasawa}, {Jonker}, {Kaastra}, {Kara}, {Karas}, {Kastner}, {King}, {Kosenko}, {Koutroumpa}, {Kraft}, {Kreykenbohm}, {Lallement}, {Lanzuisi}, {Lee}, {Lemoine-Goumard}, {Lobban}, {Lodato}, {Lovisari}, {Lotti}, {McCharthy}, {McNamara}, {Maggio}, {Maiolino}, {De Marco}, {de Martino}, {Mateos}, {Matt}, {Maughan}, {Mazzotta}, {Mendez}, {Merloni}, {Micela}, {Miceli}, {Mignani}, {Miller}, {Miniutti}, {Molendi}, {Montez}, {Moretti}, {Motch}, {Naz{\'e}}, {Nevalainen}, {Nicastro}, {Nulsen}, {Ohashi}, {O'Brien}, {Osborne}, {Oskinova}, {Pacaud}, {Paerels}, {Page}, {Papadakis}, {Pareschi}, {Petre}, {Petrucci}, {Piconcelli}, {Pillitteri}, {Pinto}, {de Plaa}, {Pointecouteau}, {Ponman}, {Ponti}, {Porquet}, {Pounds}, {Pratt}, {Predehl}, {Proga}, {Psaltis}, {Rafferty}, {Ramos-Ceja}, {Ranalli}, {Rasia}, {Rau}, {Rauw}, {Rea}, {Read}, {Reeves},
  {Reiprich}, {Renaud}, {Reynolds}, {Risaliti}, {Rodriguez}, {Rodriguez Hidalgo}, {Roncarelli}, {Rosario}, {Rossetti}, {Rozanska}, {Rovilos}, {Salvaterra}, {Salvato}, {Di Salvo}, {Sanders}, {Sanz-Forcada}, {Schawinski}, {Schaye}, {Schwope}, {Sciortino}, {Severgnini}, {Shankar}, {Sijacki}, {Sim}, {Schmid}, {Smith}, {Steiner}, {Stelzer}, {Stewart}, {Strohmayer}, {Str{\"u}der}, {Sun}, {Takei}, {Tatischeff}, {Tiengo}, {Tombesi}, {Trinchieri}, {Tsuru}, {Ud-Doula}, {Ursino}, {Valencic}, {Vanzella}, {Vaughan}, {Vignali}, {Vink}, {Vito}, {Volonteri}, {Wang}, {Webb}, {Willingale}, {Wilms}, {Wise}, {Worrall}, {Young}, {Zampieri}, {In't Zand}, {Zane}, {Zezas}, {Zhang}, \& {Zhuravleva}}]{Nandra+2013}
{Nandra}, K., {Barret}, D., {Barcons}, X., {et~al.} 2013, arXiv e-prints, arXiv:1306.2307, \dodoi{10.48550/arXiv.1306.2307}

\bibitem[{{Novikov} \& {Thorne}(1973)}]{Novikov+1973}
{Novikov}, I.~D., \& {Thorne}, K.~S. 1973, in Black Holes (Les Astres Occlus), 343--450

\bibitem[{{Ogle} {et~al.}(2003){Ogle}, {Brookings}, {Canizares}, {Lee}, \& {Marshall}}]{Ogle+2003}
{Ogle}, P.~M., {Brookings}, T., {Canizares}, C.~R., {Lee}, J.~C., \& {Marshall}, H.~L. 2003, \aap, 402, 849, \dodoi{10.1051/0004-6361:20021647}

\bibitem[{{Ogorzalek} {et~al.}(2022){Ogorzalek}, {King}, {Allen}, {Raymond}, \& {Wilkins}}]{Ogorzalek+2022}
{Ogorzalek}, A., {King}, A.~L., {Allen}, S.~W., {Raymond}, J.~C., \& {Wilkins}, D.~R. 2022, \mnras, 516, 5027, \dodoi{10.1093/mnras/stac2389}

\bibitem[{{Oh} {et~al.}(2018){Oh}, {Koss}, {Markwardt}, {Schawinski}, {Baumgartner}, {Barthelmy}, {Cenko}, {Gehrels}, {Mushotzky}, {Petulante}, {Ricci}, {Lien}, \& {Trakhtenbrot}}]{BAT+105+month+survay}
{Oh}, K., {Koss}, M., {Markwardt}, C.~B., {et~al.} 2018, \apjs, 235, 4, \dodoi{10.3847/1538-4365/aaa7fd}

\bibitem[{{Padovani} {et~al.}(2024){Padovani}, {Resconi}, {Ajello}, {Bellenghi}, {Bianchi}, {Blasi}, {Huang}, {Gabici}, {G{\'a}mez Rosas}, {Niederhausen}, {Peretti}, {Eichmann}, {Guetta}, {Lamastra}, \& {Shimizu}}]{Padovani+2024}
{Padovani}, P., {Resconi}, E., {Ajello}, M., {et~al.} 2024, Nature Astronomy, 8, 1077, \dodoi{10.1038/s41550-024-02339-z}

\bibitem[{Paggi {et~al.}(2012)Paggi, Wang, Fabbiano, Elvis, \& Karovska}]{Paggi+2012}
Paggi, A., Wang, J., Fabbiano, G., Elvis, M., \& Karovska, M. 2012, The Astrophysical Journal, 756, 39, \dodoi{10.1088/0004-637X/756/1/39}

\bibitem[{Parker {et~al.}(2013)Parker, Marinucci, Brenneman, Fabian, Kara, Matt, \& Walton}]{Parker+2013}
Parker, M.~L., Marinucci, A., Brenneman, L., {et~al.} 2013, Monthly Notices of the Royal Astronomical Society, 437, 721, \dodoi{10.1093/mnras/stt1925}

\bibitem[{{Porquet} \& {Dubau}(2000)}]{Porquet+2000}
{Porquet}, D., \& {Dubau}, J. 2000, \aaps, 143, 495, \dodoi{10.1051/aas:2000192}

\bibitem[{{Pouliasis} {et~al.}(2020){Pouliasis}, {Mountrichas}, {Georgantopoulos}, {Ruiz}, {Yang}, \& {Bonanos}}]{Pouliasis+2020}
{Pouliasis}, E., {Mountrichas}, G., {Georgantopoulos}, I., {et~al.} 2020, \mnras, 495, 1853, \dodoi{10.1093/mnras/staa1263}

\bibitem[{{Robert} \& {Casella}(2008)}]{Robert+2008}
{Robert}, C., \& {Casella}, G. 2008, arXiv e-prints, arXiv:0808.2902, \dodoi{10.48550/arXiv.0808.2902}

\bibitem[{{Rodr{\'\i}guez-Ardila} {et~al.}(2017){Rodr{\'\i}guez-Ardila}, {Mason}, {Martins}, {Ramos Almeida}, {Riffel}, {Riffel}, {Lira}, {Gonz{\'a}lez Mart{\'\i}n}, {Dametto}, {Flohic}, {Ho}, {Ruschel-Dutra}, {Thanjavur}, {Colina}, {McDermid}, {Perlman}, \& {Winge}}]{Rod+2017}
{Rodr{\'\i}guez-Ardila}, A., {Mason}, R.~E., {Martins}, L., {et~al.} 2017, \mnras, 465, 906, \dodoi{10.1093/mnras/stw2642}

\bibitem[{{Salvatore} {et~al.}(2024){Salvatore}, {Eichmann}, {Rodrigues}, {Dettmar}, \& {Becker Tjus}}]{Salvatore+2024}
{Salvatore}, S., {Eichmann}, B., {Rodrigues}, X., {Dettmar}, R.~J., \& {Becker Tjus}, J. 2024, \aap, 687, A139, \dodoi{10.1051/0004-6361/202348447}

\bibitem[{Schwarz(1978)}]{Schwarz+1978}
Schwarz, G. 1978, The Annals of Statistics, 6, 461 , \dodoi{10.1214/aos/1176344136}

\bibitem[{{Takahashi} {et~al.}(2016){Takahashi}, {Kokubun}, {Mitsuda}, {Kelley}, {Ohashi}, {Aharonian}, {Akamatsu}, {Akimoto}, {Allen}, {Anabuki}, {Angelini}, {Arnaud}, {Asai}, {Audard}, {Awaki}, {Axelsson}, {Azzarello}, {Baluta}, {Bamba}, {Bando}, {Bautz}, {Bialas}, {Blandford}, {Boyce}, {Brenneman}, {Brown}, {Bulbul}, {Cackett}, {Canavan}, {Chernyakova}, {Chiao}, {Coppi}, {Costantini}, {de Plaa}, {den Herder}, {DiPirro}, {Done}, {Dotani}, {Doty}, {Ebisawa}, {Eckart}, {Enoto}, {Ezoe}, {Fabian}, {Ferrigno}, {Foster}, {Fujimoto}, {Fukazawa}, {Furuzawa}, {Galeazzi}, {Gallo}, {Gandhi}, {Gilmore}, {Giustini}, {Goldwurm}, {Gu}, {Guainazzi}, {Haas}, {Haba}, {Hagino}, {Hamaguchi}, {Harayama}, {Harrus}, {Hatsukade}, {Hayashi}, {Hayashi}, {Hayashida}, {Hiraga}, {Hirose}, {Hornschemeier}, {Hoshino}, {Hughes}, {Ichinohe}, {Iizuka}, {Inoue}, {Inoue}, {Ishibashi}, {Ishida}, {Ishikawa}, {Ishimura}, {Ishisaki}, {Itoh}, {Iwata}, {Iyomoto}, {Jewell}, {Kaastra}, {Kallman}, {Kamae}, {Kara}, {Kataoka}, {Katsuda}, {Katsuta},
  {Kawaharada}, {Kawai}, {Kawano}, {Kawasaki}, {Khangulyan}, {Kilbourne}, {Kimball}, {King}, {Kitaguchi}, {Kitamoto}, {Kitayama}, {Kohmura}, {Kosaka}, {Koujelev}, {Koyama}, {Koyama}, {Kretschmar}, {Krimm}, {Kubota}, {Kunieda}, {Laurent}, {Lebrun}, {Lee}, {Leutenegger}, {Limousin}, {Loewenstein}, {Long}, {Lumb}, {Madejski}, {Maeda}, {Maier}, {Makishima}, {Markevitch}, {Masters}, {Matsumoto}, {Matsushita}, {McCammon}, {McGuinness}, {McNamara}, {Mehdipour}, {Miko}, {Miller}, {Miller}, {Mineshige}, {Minesugi}, {Mitsuishi}, {Miyazawa}, {Mizuno}, {Mori}, {Mori}, {Moroso}, {Moseley}, {Muench}, {Mukai}, {Murakami}, {Murakami}, {Mushotzky}, {Nagano}, {Nagino}, {Nakagawa}, {Nakajima}, {Nakamori}, {Nakano}, {Nakashima}, {Nakazawa}, {Namba}, {Natsukari}, {Nishioka}, {Nobukawa}, {Nobukawa}, {Noda}, {Nomachi}, {O'Dell}, {Odaka}, {Ogawa}, {Ogawa}, {Ogi}, {Ohno}, {Ohta}, {Okajima}, {Okamoto}, {Okazaki}, {Ota}, {Ozaki}, {Paerels}, {Paltani}, {Parmar}, {Petre}, {Pinto}, {Pohl}, {Pontius}, {Porter}, {Pottschmidt}, {Ramsey},
  {Reynolds}, {Russell}, {Safi-Harb}, {Saito}, {Sakai}, {Sakai}, {Sameshima}, {Sasaki}, {Sato}, {Sato}, {Sato}, {Sato}, {Sawada}, {Schartel}, {Serlemitsos}, {Seta}, {Shibano}, {Shida}, {Shidatsu}, {Shimada}, {Shinozaki}, {Shirron}, {Simionescu}, {Simmons}, {Smith}, {Sneiderman}, {Soong}, {Stawarz}, {Sugawara}, {Sugita}, {Sugita}, {Szymkowiak}, {Tajima}, {Takahashi}, {Takeda}, {Takei}, {Tamagawa}, {Tamura}, {Tamura}, {Tanaka}, {Tanaka}, {Tanaka}, {Tashiro}, {Tawara}, {Terada}, {Terashima}, {Tombesi}, {Tomida}, {Tsuboi}, {Tsujimoto}, {Tsunemi}, {Tsuru}, {Uchida}, {Uchiyama}, {Uchiyama}, {Ueda}, {Ueda}, {Ueno}, {Uno}, {Urry}, {Ursino}, {de Vries}, {Wada}, {Watanabe}, {Watanabe}, {Werner}, {Wik}, {Wilkins}, {Williams}, {Yamada}, {Yamada}, {Yamaguchi}, {Yamaoka}, {Yamasaki}, {Yamauchi}, {Yamauchi}, {Yaqoob}, {Yatsu}, {Yonetoku}, {Yoshida}, {Yuasa}, {Zhuravleva}, \& {Zoghbi}}]{Takahashi+2016}
{Takahashi}, T., {Kokubun}, M., {Mitsuda}, K., {et~al.} 2016, in Society of Photo-Optical Instrumentation Engineers (SPIE) Conference Series, Vol. 9905, \procspie, 99050U, \dodoi{10.1117/12.2232379}

\bibitem[{{Tarter} {et~al.}(1969){Tarter}, {Tucker}, \& {Salpeter}}]{Tarter+1969}
{Tarter}, C.~B., {Tucker}, W.~H., \& {Salpeter}, E.~E. 1969, \apj, 156, 943, \dodoi{10.1086/150026}

\bibitem[{{Tashiro} {et~al.}(2020){Tashiro}, {Maejima}, {Toda}, {Kelley}, {Reichenthal}, {Hartz}, {Petre}, {Williams}, {Guainazzi}, {Costantini}, {Fujimoto}, {Hayashida}, {Henegar-Leon}, {Holland}, {Ishisaki}, {Kilbourne}, {Loewenstein}, {Matsushita}, {Mori}, {Okajima}, {Porter}, {Sneiderman}, {Takei}, {Terada}, {Tomida}, {Yamaguchi}, {Watanabe}, {Akamatsu}, {Arai}, {Audard}, {Awaki}, {Babyk}, {Bamba}, {Bando}, {Behar}, {Bialas}, {Boissay-Malaquin}, {Brenneman}, {Brown}, {Canavan}, {Chiao}, {Comber}, {Corrales}, {Cumbee}, {de Vries}, {den Herder}, {Dercksen}, {Diaz-Trigo}, {DiPirro}, {Done}, {Dotani}, {Ebisawa}, {Eckart}, {Eckert}, {Eguchi}, {Enoto}, {Ezoe}, {Ferrigno}, {Fujita}, {Fukazawa}, {Furuzawa}, {Gallo}, {Gorter}, {Grim}, {Gu}, {Hagino}, {Hamaguchi}, {Hatsukade}, {Hawthorn}, {Hayashi}, {Hell}, {Hiraga}, {Hodges-Kluck}, {Horiuchi}, {Hornschemeier}, {Hoshino}, {Ichinohe}, {Iga}, {Iizuka}, {Ishida}, {Ishihama}, {Ishikawa}, {Ishimura}, {Jaffe}, {Kaastra}, {Kallman}, {Kara}, {Katsuda}, {Kenyon}, {Kimball},
  {Kitaguchi}, {Kitamoto}, {Kobayashi}, {Kobayashi}, {Kohmura}, {Kubota}, {Leutenegger}, {Li}, {Lockard}, {Maeda}, {Markevitch}, {Martz}, {Matsumoto}, {Matsuzaki}, {McCammon}, {McLaughlin}, {McNamara}, {Miko}, {Miller}, {Miller}, {Minesugi}, {Mitani}, {Mitsuishi}, {Mizumoto}, {Mizuno}, {Mukai}, {Murakami}, {Mushotzky}, {Nakajima}, {Nakamura}, {Nakazawa}, {Natsukari}, {Nigo}, {Nishioka}, {Nobukawa}, {Nobukawa}, {Noda}, {Odaka}, {Ogawa}, {Ohashi}, {Ohno}, {Ohta}, {Okamoto}, {Ota}, {Ozaki}, {Paltani}, {Plucinsky}, {Pottschmidt}, {Sampson}, {Sasaki}, {Sato}, {Sato}, {Sato}, {Sawada}, {Seta}, {Shibano}, {Shida}, {Shidatsu}, {Shigeto}, {Shinozaki}, {Shirron}, {Simionescu}, {Smith}, {Someya}, {Soong}, {Sugawara}, {Sugawara}, {Szymkowiak}, {Takahashi}, {Takeshima}, {Tamagawa}, {Tamura}, {Tanaka}, {Tanimoto}, {Terashima}, {Tsuboi}, {Tsujimoto}, {Tsunemi}, {Tsuru}, {Uchida}, {Uchida}, {Uchiyama}, {Ueda}, {Uno}, {Vink}, {Watanabe}, {Witthoeft}, {Wolfs}, {Yamada}, {Yamaoka}, {Yamasaki}, {Yamauchi}, {Yamauchi},
  {Yanagase}, {Yaqoob}, {Yasuda}, {Yoshida}, {Yoshioka}, \& {Zhuravleva}}]{Tashiro+2020}
{Tashiro}, M., {Maejima}, H., {Toda}, K., {et~al.} 2020, in Society of Photo-Optical Instrumentation Engineers (SPIE) Conference Series, Vol. 11444, Space Telescopes and Instrumentation 2020: Ultraviolet to Gamma Ray, ed. J.-W.~A. {den Herder}, S.~{Nikzad}, \& K.~{Nakazawa}, 1144422, \dodoi{10.1117/12.2565812}

\bibitem[{{Tomaru} {et~al.}(2023){Tomaru}, {Done}, \& {Mao}}]{Tomaru+2023}
{Tomaru}, R., {Done}, C., \& {Mao}, J. 2023, \mnras, 518, 1789, \dodoi{10.1093/mnras/stac3210}

\bibitem[{{Tombesi} {et~al.}(2013){Tombesi}, {Cappi}, {Reeves}, {Nemmen}, {Braito}, {Gaspari}, \& {Reynolds}}]{Tombesi+2013}
{Tombesi}, F., {Cappi}, M., {Reeves}, J.~N., {et~al.} 2013, \mnras, 430, 1102, \dodoi{10.1093/mnras/sts692}

\bibitem[{{Urry} \& {Padovani}(1995)}]{Urry+1995}
{Urry}, C.~M., \& {Padovani}, P. 1995, \pasp, 107, 803, \dodoi{10.1086/133630}

\bibitem[{{Wang} {et~al.}(2011){Wang}, {Fabbiano}, {Elvis}, {Risaliti}, {Mundell}, {Karovska}, \& {Zezas}}]{Wang+2011}
{Wang}, J., {Fabbiano}, G., {Elvis}, M., {et~al.} 2011, \apj, 736, 62, \dodoi{10.1088/0004-637X/736/1/62}

\bibitem[{Weisskopf {et~al.}(2000)Weisskopf, Tananbaum, Speybroeck, \& O'Dell}]{Weisskopf+2000}
Weisskopf, M.~C., Tananbaum, H.~D., Speybroeck, L. P.~V., \& O'Dell, S.~L. 2000, in X-Ray Optics, Instruments, and Missions III, ed. J.~E. Truemper \& B.~Aschenbach, Vol. 4012, International Society for Optics and Photonics (SPIE), 2 -- 16, \dodoi{10.1117/12.391545}

\bibitem[{{Wilkins} \& {Gallo}(2015)}]{Wilkins+2015B}
{Wilkins}, D.~R., \& {Gallo}, L.~C. 2015, \mnras, 449, 129, \dodoi{10.1093/mnras/stv162}

\bibitem[{Wilkins {et~al.}(2014)Wilkins, Kara, Fabian, \& Gallo}]{Wilkins+2014}
Wilkins, D.~R., Kara, E., Fabian, A.~C., \& Gallo, L.~C. 2014, Monthly Notices of the Royal Astronomical Society, 443, 2746, \dodoi{10.1093/mnras/stu1273}

\bibitem[{{Willingale} {et~al.}(2013){Willingale}, {Starling}, {Beardmore}, {Tanvir}, \& {O'Brien}}]{Willingale+2013}
{Willingale}, R., {Starling}, R.~L.~C., {Beardmore}, A.~P., {Tanvir}, N.~R., \& {O'Brien}, P.~T. 2013, \mnras, 431, 394, \dodoi{10.1093/mnras/stt175}

\bibitem[{{XRISM Collaboration}(2024)}]{XRISM+2024-NGC4151}
{XRISM Collaboration}. 2024, arXiv e-prints, arXiv:2408.14300, \dodoi{10.48550/arXiv.2408.14300}

\bibitem[{{Xu} {et~al.}(2025){Xu}, {Gallo}, {Hagino}, {Reeves}, {Tombesi}, {Mizumoto}, {Luminari}, {Gonzalez}, {Behar}, {Boissay-Malaquin}, {Braito}, {Condo}, {Done}, {Miyamoto}, {Mizukawa}, {Odaka}, {Sato}, {Tanimoto}, {Tashiro}, {Yaqoob}, \& {Yamada}}]{Xu+2025}
{Xu}, Y., {Gallo}, L.~C., {Hagino}, K., {et~al.} 2025, arXiv e-prints, arXiv:2506.05273, \dodoi{10.48550/arXiv.2506.05273}

\bibitem[{{Zaino} {et~al.}(2020){Zaino}, {Bianchi}, {Marinucci}, {Matt}, {Bauer}, {Brandt}, {Gandhi}, {Guainazzi}, {Iwasawa}, {Puccetti}, {Ricci}, \& {Walton}}]{Zaino+2020}
{Zaino}, A., {Bianchi}, S., {Marinucci}, A., {et~al.} 2020, \mnras, 492, 3872, \dodoi{10.1093/mnras/staa107}

\end{thebibliography}
\bibliographystyle{aasjournal}
\appendix

\section{Fe {\sc xxii} Excitation and Emission Processes in an AGN Radiation Field}
\label{apen}
\begin{figure*}[hbt!]
    \centering
    \includegraphics[width=1\linewidth]{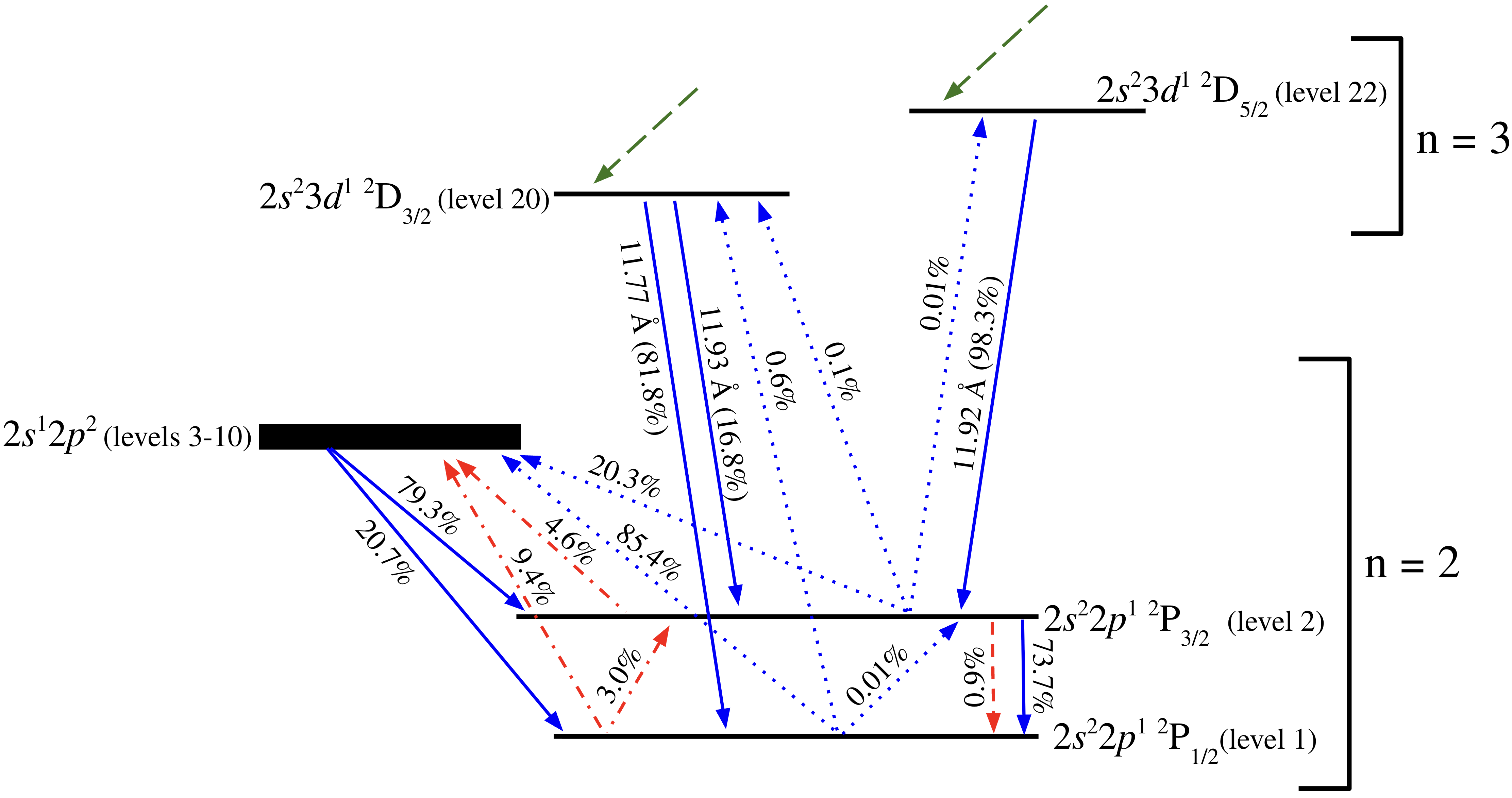}
    \caption{Energy level diagram for the \fe\ doublet in photoionized equilibrium under best fit conditions. Each level is labelled with its electron configuration, term symbol, and energy level (e.g., the ground state is level 1, the first excited state is level 2, and so on). The $2s^12p^2$ manifold contains levels $3-10$, and the term symbols are neglected. They are treated as one energy state for simplistically. Blue solid lines indicate radiative decay, blue dotted lines indicate absorption. Red dot dashed lines indicate the collisional excitation, and the red dashed line shows collisional de-excitation. The green dashed line is the combination of radiative recombination and cascades. 
    The percentages listed provide information about the location and method by which an electron will transition from each level. The excitation energy between level 1 and level 2 is $14.7$\,eV (845.55\,\AA). The excitation energies from level 1 to levels $3-10$ range between $\sim50-125$\,eV ($\sim250-100$\,\AA). }
    \label{fig:lvlD}
\end{figure*}

As discussed in the text, the \fe\ density diagnostic has been commonly used to determine the density of material surrounding magnetic cataclysmic variable stars under collisionally ionized equilibrium (CIE) conditions \citep[e.g.][]{Mauche+2003, Mauche+2004, Chen+2004, Liang+2008}. The density diagnostic is dependant on the population of the first excited state ($2s^22p^1\,^2$P$_{3/2}$) compared to the ground state ($2s^22p^1\,^2$P$_{1/2}$). The transitions that form our diagnostic line are $2s^2 3d^1 \,{ }^2{\rm D}_{3/2}\,\rightarrow \, 2s^2 2p^1 \, { }^2{\rm P}_{1/2}$ and $2s^2 3d^1 \,{ }^2{\rm D}_{5/2}\,\rightarrow \,2s^2 2p^1 \, { }^2{\rm P}_{3/2}$. These are bound-bound transitions between the nineteenth excited state and the ground state, and between the twenty-first excited state and the first excited state, respectively. For the rest of this discussion, we will refer to the ground state as level 1, the first excited state as level 2, the nineteenth excited state as level 20 and the twenty-first excited state as level 22. These are the labels used by {\sc spex}. Levels three to ten are also important for this discussion, as they form the $2s^12p^2$ manifold, and they will be labelled as levels $3-10$. Figure \ref{fig:lvlD} shows the energy level diagram with the electron configuration, term symbols and labels for the levels of interest. 

The diagnostic curves in the literature assume that the predominant excitation mechanism is collisional. Ions in the ground state or the first excited state will collisionally excite into either level 2 (ground state only), levels $3-10$, level 20 or level 22. Most collisional excitations will place the electrons in levels $3-10$, which will quickly decay back to either level 1 or level 2. 

To produce the observed emission lines, we require radiative decays from levels 20 and 22. 
If an electron is excited to level 20, it will most likely decay to the ground state, producing the 11.77\,\AA\ emission line. It can also decay to level 2 about $15\,$percent of the time, producing the 11.93\,\AA\ emission line. 
If the electron is excited to level 22, it will decay to level 2, producing the 11.92\,\AA\ emission line. No radiative transition can occur between level 22 and the ground state. 

An electron in level 2 is more likely to excite to level 22 than to level 20. An electron in the ground state is more likely to excite to level 20 than to level 22.  Thus, the ratio of the emission from the 11.77\,\AA\ and 11.92\,\AA\ gives an indication of the respective populations of the ground and first excited state. 

The lifetime of the level  $2\rightarrow 1$ transition is $\sim 0.7$s \citep{NIST+2024}. In the low-density limit, excited ions with electrons in level 2 will have sufficient time to decay before a second interaction can take place to excite to levels 20 or 22. In this limit, the population of electrons in level 2 will remain small, and the 11.92\,\AA\ line will be weak. As the density increases, more electrons are collisionally excited to all levels. The majority of electrons that are excited to levels $3-10$ will decay to level 2, significantly populating this level. Thus, the population of level 2 electrons will build and excitation to level 22 will exceed excitation to level 20. Thus, the 11.92\,\AA\ line will be strong. 

In a radiation field, under photoionized equilibrium (PIE), the fundamental mechanism of the diagnostic ratio is the same. The population of levels 20 and 22 depend on the populations of the ground and first excited states. Figure \ref{fig:lvlD} shows an energy level diagram for the \fe\ density diagnostic for the best-fit High ionization component. 
Absorption is shown with blue dotted lines, and emission is shown with blue solid lines. Collisional excitations are shown with red dot-dashed lines, while collisional de-excitation is shown with a red dashed line. The green, long-dashed arrows represent both radiative recombination and cascades from higher energy levels. The number of electrons flowing (via collisional or radiative processes) relative to the total number of electrons leaving a given level is given as percentages in Figure \ref{fig:lvlD}. The information has been extracted using the {\sc spex} commands {\sc ascdump lev} and {\sc ascdump pop}.

While collisional excitations between level 1, level 2 and levels $3-10$ are not negligible, they are, however, less important than the radiative transitions. Perhaps most important are the numerous radiative transitions that promote electrons to levels $3-10$. In the low-density limit, the electrons in level 1 are promoted to levels $3-10$ by radiative, and to a lesser extent, collisional transitions. Electrons in level 2 almost exclusively decay to level 1 before any secondary excitation can occur. This results in a very limited population in level 22, as this cannot be radiatively populated from ions in the ground state. Therefore, in the low-density limit, the only method by which electrons can reach level 22 is through radiative recombination and cascades from higher energy levels. In this case, the 11.92\,\AA\ line is weak and is overshadowed by the 11.93\,\AA\ line originating from the $20\,\rightarrow\,2$ transition

In Figure \ref{fig:lvlD}, where the density is reaching critical, we can see that the population of level 2 is beginning to build up. The primary method for electrons leaving this level is still radiative decay to level 1. However, a significant fraction of electrons are excited to levels $3-10$, level 20 and level 22. This allows the population of level 22 to increase and allows for direct radiative transition to level 22. The 11.92\,\AA\ emission line gains strength. 

In the high-density case, there are very limited radiative transitions between level 1 and level 2. Electron exchanges between these levels are more likely to be collisional in nature. The fraction of radiative transitions between level 2 and level 22 increases, further strengthening the 11.92\,\AA\ emission line. 

As noted in the text, the critical density of the diagnostic curves produced with PIE plasmas is significantly smaller (up to three orders of magnitude) than that seen in the literature. We theorize that this effect results from the large number of continuum photons that are absorbed from levels 1 and 2, into levels $3-10$. The ground-state excitation energy for levels $3-10$ ranges from $\sim50-125\,$eV, and the AGN continuum can readily supply these photons. Once ions are excited to levels $3-10$, they will preferentially decay to level 2. This process, in addition to the collisional excitations that occur as the density increases, will populate level 2. With the radiative transitions aiding the collisional excitations, the population of level 2 will increase to the point where excitations between level 2 and level 22 are common at a lower critical density. 

To explicitly demonstrate how a radiation field affects the \fe\ density sensitive doublet, and the resultant diagnostic curves, we create a non-equilibrium hybrid plasma by modifying  {\sc pion} to have a fixed of 1\,keV\footnote{We do this by setting the parameter {\sc tmod}\,$=1$}. While this collisionally ionized {\sc pion} ({\sc CI-pion}) plasma is not in equilibrium, we can extract information about how an AGN radiation field affects these density-sensitive emission lines when collisional excitations are non-negligible. 

\begin{figure}
    \centering
    \includegraphics[width=\linewidth]{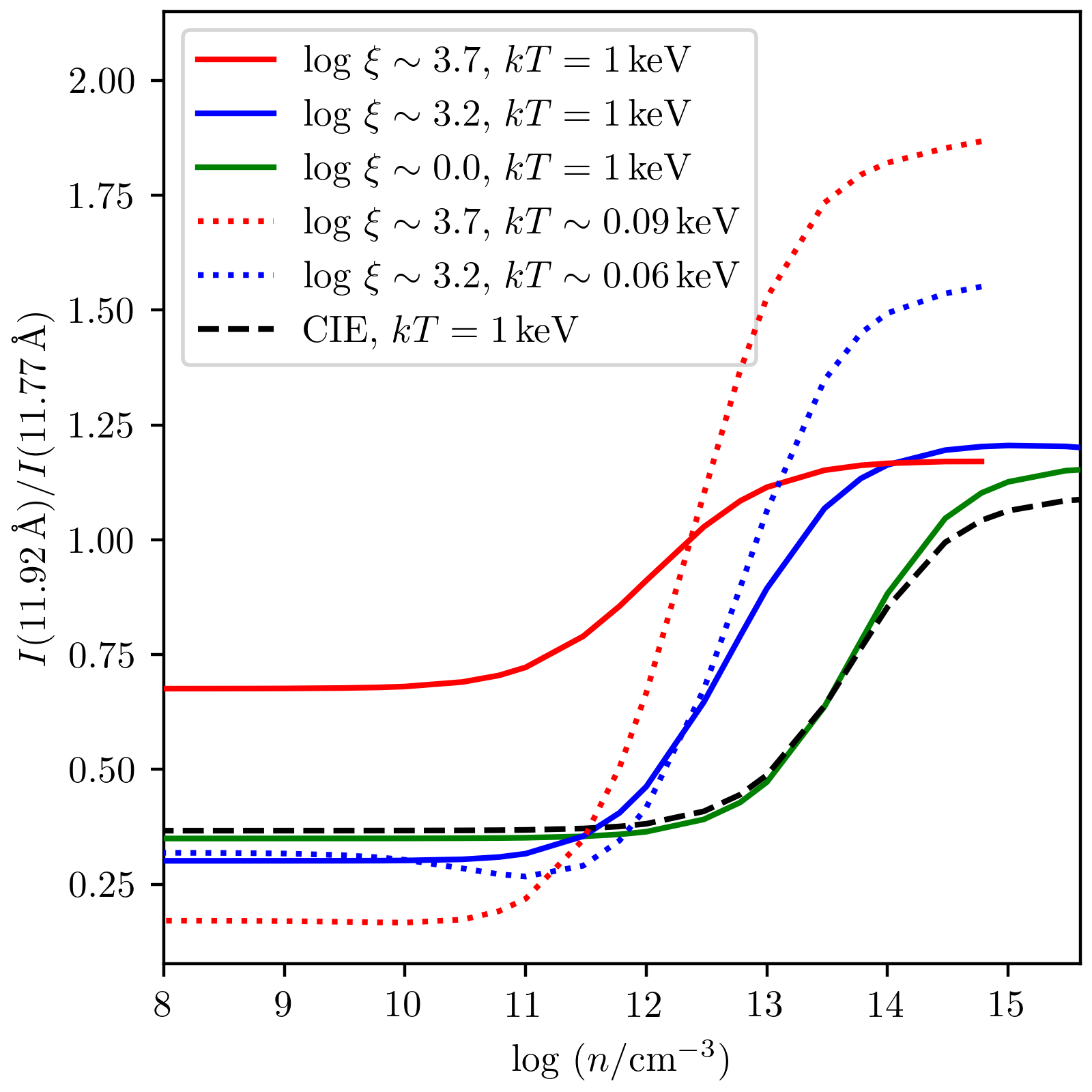}
    \caption{Diagnostic curves for the CI-{\sc pion} model. The solid red, blue, and green curves indicate that for a plasma with $kT=1\,{\rm keV}$, the critical density of the diagnostic lines increases as ionization decreases. The dotted red and blue curves are included to illustrate what the curves would look like without temperature manipulation (i.e. in a purely photoionized scenario). The black dashed line is the diagnostic curve produced by a plasma in CIE with a temperature of 1\,keV. }
    \label{fig:ci-pion}
\end{figure}

We generate diagnostic curves similar to those produced in Section \ref{sec:diagnostic}, with the temperature fixed at 1\,keV for log\,$(\xi/\ergpcms)=0.0,3.2,$ and 3.7. These curves are shown by the solid curves in Figure \ref{fig:ci-pion}. For comparison, we include the purely photoionized curves for log\,$(\xi/\ergpcms) = 3.2$ and 3.7. They are shown as blue and red dotted curves, respectively. The critical densities in these purely photoionized cases are comparable to those found in the hybrid plasma for log\,$(\xi/\ergpcms) = 3.2$ and 3.7. The range of values of $I(11.92$\,\AA$)/I(11.77$\,\AA$)$ is significantly different in the hybrid plasma. 

No purely photoionized diagnostic curve is produced for log\,$(\xi/\ergpcms)=0.0$ as this ionization is insufficient to produce \fe\ ions.  We infer that the behaviour of the {\sc CI-pion} with log\,$(\xi/\ergpcms) = 0.0$ is a good approximation of a CIE plasma. This is confirmed by the diagnostic curve produced using {\sc cie}\footnote{Using  SPEXACT v3.08.01} \citep{Kaastra+1996},  shown with the black dashed line. It is well matched to the {\sc CI-pion} with log\,$(\xi/\ergpcms)=0.0$, in both critical density and values of $I(11.92$\,\AA$)/I(11.77$\,\AA$)$. These two curves are remarkably similar to each other and to the literature values presented by \cite{Mauche+2003}.

With {\sc CI-pion}, we can probe the effect of lower ionization on the diagnostic doublet. As the ionization of the plasma decreases, the critical density increases. This behaviour is also observed in Figure \ref{fig:bigFig}, but is shown more clearly in Figure \ref{fig:ci-pion}. The critical density is increasing due to a reduction in the number of continuum photons that can excite to levels $3-10$. Instead, as the ionization decreases, the plasma requires more collisional excitation to populate level 2, and strengthen the 11.92\,\AA\ emission line. The population inversion then occurs at higher densities, resulting in a higher critical density for the diagnostic. 

Thus, we are confident in concluding that the effect of an AGN radiation field is to photoexcite ions to level 2 via the $2s^12p^2$ manifold. This thereby reduces the critical density of this diagnostic ratio. 

\end{document}